\documentstyle[psfig]{mn}

\begin{document}

\journal{To Appear in MNRAS}

\title[X-ray spectral analysis of elliptical galaxies] {X-ray spectral
analysis of elliptical galaxies from {\em ASCA}: \\ the Fe abundance
in a multi-phase medium } \author[D.~A. Buote and A.~C. Fabian]{David
A. Buote and A. C. Fabian \\ Institute of Astronomy, Madingley
Road, Cambridge CB3 0HA}

\maketitle

\begin{abstract}

We present spectral analysis of {\sl ASCA} data of 17 elliptical
and 3 lenticular galaxies most of which have high $L_{\rm x}/L_{\rm B}$.
Single-temperature models (MEKAL and Raymond-Smith) give unacceptable
fits $(\chi^2_{\rm red}>1.5)$ in most cases and, in agreement with
previous studies, give very sub-solar abundances, $\langle Z\rangle =
0.19\pm 0.12$ $Z_{\sun}$ (MEKAL).  The spectra for approximately half
the sample are better fit with a cooling-flow model which in three
cases gives a substantially better fit.  The abundances derived from
the cooling-flow model are also significantly larger, $\langle
Z\rangle = 0.6\pm 0.5$ $Z_{\sun}$. We empirically tested the
reliability of the plasma codes in the Fe-L region and found no
evidence for serious problems with the determined temperatures and
abundances.

Two-temperature models give substantially better fits that are
formally acceptable ($\chi^2_{\rm red}\sim 1.0$) in all but a few
cases. The highest S/N galaxies (which also have highest $L_{\rm
x}/L_{\rm B}$) have fitted temperatures $<2$ keV for both components
consistent with each being distinct phases of hot gas. The lowest S/N
galaxies (which also have lowest $L_{\rm x}/L_{\rm B}$) generally have
a hot component with temperature, $T_{\rm H}\ga 5$ keV, which is
consistent with emission from discrete sources.  (We discuss the
origin of these two components from analysis of $L_{\rm x}-L_{\rm B}$
and $L_{\rm x}/L_{\rm B}$.)  The abundances of these two-temperature
models are approximately solar, $\langle Z\rangle = 0.9\pm 0.7$
$Z_{\sun}$ (MEKAL), consistent with a recent multi-phase model for the
evolution of hot gas in ellipticals. Finally, for several galaxies we
find evidence for absorption in excess of the Galactic value and
discuss its implications.

\end{abstract}

 
\section{Introduction}
\label{intro}

The metallicity of the hot X-ray emitting gas in ellipticals is a
sensitive diagnostic of their supernova history and thus serves as a
powerful discriminant for models of the enrichment of the hot gas
(e.g. Loewenstein \& Mathews 1991; Ciotti et al. 1991). Previous {\it
ROSAT} and {\sl ASCA} studies have found the metallicities of
ellipticals to be very sub-solar in contradiction to the standard
enrichment models (e.g. Sarazin 1997).  This discrepancy has led some
authors \cite{arimoto} to question the reliability of the standard
plasma codes, in particular their modeling of the Fe-L emission lines.

However, the abundances derived from the plasma codes also depend on
the temperature structure of the gas.  Because the cooling times are
short ($\sim 10^9$ yr) in ellipticals (e.g. Thomas et al. 1986) the
hot gas in elliptical galaxies may consist of a continuum of hot and
cold phases similar to that expected for galaxy clusters (Nulsen 1986;
Thomas, Fabian, \& Nulsen 1987).  Fitting single-temperature models to
intrinsically two-temperature spectra can result in a significant
underestimate of the metallicity \cite{BC94}.  

Matsumoto et al. \shortcite{mat} have obtained very sub-solar
abundances from fitting two-component models to {\sl ASCA} SIS and GIS
data of a sample of 12 ellipticals, but they restricted the form of
the high-temperature component to have the same temperature for each
galaxy.  That is, Matsumoto et al. combined the data at high energies,
$E\sim 4-10$ keV, for the 12 galaxies and found that the composite
spectrum is not tightly constrained but equally well described by
either a thermal bremsstrahlung model with $T\sim 12$ keV or a
power-law model with photon index 1.8.  These models, which are
thought to describe the emission from discrete sources, are then
assumed to be similar for every galaxy and to apply over the full {\it
ASCA} energy band, $E=0.4-10$ keV.  However, these models need not be
valid representations for every galaxy, especially for energies below
4 keV where they are only extrapolations.

Hence, there is a need for a study of a larger sample of ellipticals
with fewer modeling assumptions. Moreover, the calibration status of
{\sl ASCA} has changed significantly over the past year and thus the
results of previous studies need to be verified.  We have assembled a
sample of 20 early-type galaxies (17 E, 3 S0 -- see Table
\ref{tab.prop}) with high $L_{\rm x}/L_{\rm B}$ values as listed in
Kim, Fabbiano, \& Trinchieri \shortcite{kft} to insure that the X-ray
emission is mostly due to hot gas. These galaxies have archival {\it
ASCA} observations of good quality for most of the sample. We place a
premium on the signal-to-noise ratio, S/N, and thus the extraction
radii for the spectra are in many cases smaller than in previous
studies. This attention to S/N is vital in order to achieve the global
$\chi^2$ minima with XSPEC for our models with many fitting
parameters. We fit as many temperature components as are demanded by
the data to achieve a good fit without any restrictions on the
temperatures. We also examine the reliability of the plasma codes in
the Fe-L spectral band directly from the sample.

The organization of the paper is as follows. In \S \ref{obs} we
describe the details of the observations and data analysis. The
results of the single-temperature fits and the empirical test of the
plasma codes is given in \S \ref{1t}. We discuss the two-temperature
(and three-temperature) fits in \S \ref{2t}. In \S \ref{nh} we discuss
the evidence and implications for excess absorption in several of the
galaxies. Finally, we summarize and give our conclusions in \S
\ref{conc} .

\section{Observations and data reduction}
\label{obs}

\begin{table}
\caption{SIS vs GIS}
\label{tab.sis}
\begin{tabular}{lrrrr}
Energy &  \multicolumn{2}{c}{Area (cm$^2$)} &
\multicolumn{2}{c}{$\Delta \rm E$ (eV)} \\
(keV) & SIS & GIS & SIS & GIS\\

0.75 & 60 & 6 & 70 & $\sim 200$\\
1.00 & 140 & 50 & 75 & $\sim 200$\\
1.25 & 175 & 100 & 80 & 225\\
1.50 & 200 & 130 & 85 & 240\\
2.00 & 150 & 160 & 90 & 250\\
5.00 & 130 & 120 & 150 & 400\\

\end{tabular}

\medskip

Comparison of the effective area and energy resolution of the SIS and
GIS. These values are approximate as they were obtained from visual
inspection of Figures 8.4a, 8.4b, and 9.5 of the {\sl ASCA} Technical
Description (Appendix E, AN 97-OSS-02). The spectral resolution of the
SIS refers to February 1994.

\end{table}

\begin{table*}
\begin{minipage}{122mm}
\caption{Galaxy Properties}
\label{tab.prop}
\begin{tabular}{llclcccc}
Name & Type & $z$ & $B_{\rm T}^0$ & $\log_{10} L_{\rm B}$ & $\sigma_0$
     & $N_{\rm H}$ & Group\\ 
     &      &     &       & erg s$^{-1}$ &  (km s$^{-1}$)    &
($10^{21}$cm$^{-2}$)\\  
NGC 499  & S$0^{-}$& 0.01459 & 12.64 & 43.74 & 237 & 0.53 & LGG 24\\
NGC 507  & LAR0    & 0.01642 & 11.76 & 44.19 & 366 & 0.53 & LGG 24\\
NGC 720  & E5      & 0.00572 & 11.15 & 43.47 & 247 & 0.14 & LGG 38\\
NGC 1332 & S$0^{-}$& 0.00508 & 11.29 & 43.29 & 347 & 0.22 & LGG 97\\
NGC 1399 & E1P 	& 0.00483 & 10.79 & 43.46 & 310 & 0.13 & Fornax\\
NGC 1404 & E1 	& 0.00648 & 11.06 & 43.35 & 225 & 0.13 & Fornax\\
NGC 1407 & E0 	& 0.00593 & 10.93 & 43.61 & 285 & 0.55 & LGG 100\\
NGC 3923 & E4-5 	& 0.00556 & 10.79 & 43.82 & 216 & 0.64 & LGG 253\\
NGC 4374 & E1 	& 0.00334 & 10.23 & 43.67 & 287 & 0.26 & Virgo\\
NGC 4406 & E3 	&-0.00076 & 10.02 & 43.76 & 250 & 0.26 & Virgo\\
NGC 4472 & E2 	& 0.00290 & 09.32 & 44.04 & 287 & 0.16 & Virgo\\
NGC 4636 & E$0^+$ 	& 0.00365 & 10.50 & 43.58 & 191 & 0.17 & Virgo\\
NGC 4649 & E2 	& 0.00471 & 09.83 & 43.83 & 341 & 0.24 & Virgo\\
NGC 5044 & E0 	& 0.00898 & 11.87 & 43.75 & 234 & 0.50 & LGG 338\\
NGC 5846 & E0-1 	& 0.00608 & 11.13 & 43.77 & 278 & 0.42 & LGG 393\\
NGC 6876 & SB$0^-$ & 0.01338 & 12.45 & 43.66 & 231 & 0.52 & LGG 432\\
NGC 7619 & E2 	& 0.01269 & 12.17 & 43.79 & 337 & 0.50 & LGG 473\\
NGC 7626 & E1P 	& 0.01142 & 12.17 & 43.70 & 234 & 0.50 & LGG 473\\
IC 1459 & E3 	& 0.00564 & 10.96 & 43.53 & 308 & 0.12 & LGG 466\\
IC 4296 & E0 	& 0.01255 & 11.43 & 44.12 & 323 & 0.43 & LGG 353\\
\end{tabular}

\medskip

Morphological types and redshifts are taken from RC3.  Total blue
magnitudes $(B_{\rm T}^0)$ and central velocity dispersions
$(\sigma_0)$ are taken from Faber et al. \shortcite{7s} except for NGC
1332 \cite{dalle}. The luminosities $(L_{\rm B})$ corresponding to
$B_{\rm T}^0$ are computed using the distances from Faber et
al. \shortcite{7s} as reported in Eskridge, Fabbiano, \& Kim
\shortcite{efk} for $H_0=50$ km/s/Mpc. Galactic Hydrogen column
densities $(N_{\rm H})$ are from Stark et al. \shortcite{stark}. LGG
refers to Garcia \shortcite{garcia}.

\end{minipage}
\end{table*}

The {\sl ASCA} satellite (Tanaka, Inoue, \& Holt 1994) consists of
four detectors each illuminated by a dedicated X-ray telescope.  Two
of the detectors are X-ray CCD cameras (Solid-State Imaging
Spectrometers -- SIS0 and SIS1) and two are proportional counters (Gas
Imaging Spectrometers -- GIS2 and GIS3).  Since the X-ray luminosity
of the early-type galaxies in our sample is thought to be dominated by
thermal emission with $T\sim 1$ keV
\footnote{Throughout the paper we set $k_BT\rightarrow T$.} (Forman,
Jones, \& Tucker 1985; Canizares, Fabbiano, \& Trinchieri 1987; Kim,
Fabbiano, \& Trinchieri 1992), and the effective area and energy
resolution of the SIS is substantially greater than that of the GIS
for energies $\sim 0.5-1.5$ keV (see Table \ref{tab.sis}), the SIS
data has far better S/N over the important energies for constraining
spectral models and for probing the Fe-L emission region.  

That is, the fluxes of bright early-type galaxies above 2 keV are
typically $\la 10\%$ of the $0.5-2$ keV fluxes (e.g. Table
\ref{tab.lum}) which indicates that even though the effective areas of
the SIS and GIS are comparable for energies $\sim 2-5$ keV (the SIS
being slightly {\it better} on average) the important constraints on
spectral models fitted over $\sim 0.5-5$ keV are determined almost
completely by the SIS data because of its substantial advantage in
effective area for energies below 2 keV and its superior energy
resolution over the entire {\sl ASCA} band (e.g. Buote \& Canizares
1997). Since the SIS data dominates the constraints on spectral models
of early-type galaxies, any small differences in the model parameters
obtained from jointly fitting SIS and GIS data of early-type galaxies
may be due to mismatches in calibration between the two instruments;
e.g. the gain calibration of the GIS \cite{idesawa}. Hence, we focus
on the SIS data, although throughout the paper we illustrate pertinent
results using the GIS data for representative galaxies.  The reduction
of the SIS and GIS data was performed using the standard XSELECT,
FTOOLS, and IRAF-PROS software.

We list details of the observations in Table \ref{tab.obs}.  Because
the SIS0 and SIS1 (and GIS2 and GIS3) have different energy responses
and gains their data must be fitted separately. The same applies for
different observations of the same galaxy since the energy response
and gain vary significantly with time. As a result, for a galaxy with
multiple observations, if one of the sequences has a substantially
smaller exposure time than the others, it will not contribute
significantly to the spectral fits. (And any residual differences
caused by incorporating the lower S/N observation may just be
systematics owing to background subtraction or calibration.) Moreover,
the aperture size used to extract the spectrum for the lower S/N
sequence will be necessarily smaller than the others which could lead
to inconsistencies. (We describe our determination of aperture sizes
below.) Hence, for a few galaxies with multiple observations (NGC 507,
NGC 4374, NGC 5044) we ignored these sequences with much worse S/N.

Similarly, for a given observation the SIS0 data has higher S/N than
the SIS1 data because the screening criteria for events is much more
severe for the SIS1. If a great disparity in S/N existed between the
two detectors then we analyzed only the high S/N observation (NGC
4649, NGC 6876). For NGC 507 we excluded the SIS1 data because of
anomalies that we discuss in \S \ref{fit}.

For each galaxy we analyzed the data generated by the standard
processing software that were filtered with the default screening
criteria (see The ASCA Data Reduction Guide 1997). We further screened
the data by excluding time intervals of high background from analysis
of the light curves. For NGC 720 and NGC 1332 we used the events
screened according to the criteria in Buote \& Canizares
\shortcite{BC97} modified by the light-curve analysis. 

The final filtered events were then extracted from a circular region
positioned by eye on the galaxy center. A circular region was chosen
because the standard software which incorporates information on
effective area and corrects the spectra for vignetting (i.e. FTOOLS
task {\it ascaarf}) currently performs best on circular
regions. However, for NGC 4406 we used a square region because the
X-ray emission is highly asymmetrical which offsets the benefits of a
circular region in {\it ascaarf}.

Our guiding principle for choosing the radius of the circle for a
particular galaxy was to maximize the S/N of the background-subtracted
counts within the $\sim 0.5-5$ keV energy band.  We modified the $S/N$
criterion according to other considerations. We tried to limit as much
as possible the circular aperture to a single CCD because the
responses vary from chip to chip. (For the cases where the source
counts extended over more than one chip we followed the standard
procedure and averaged the response matrices according to the number
of source counts on each chip. The gaps between chips were excluded.)
For the galaxies having a nearby bright source (e.g. NGC 1399 and NGC
1404; NGC 499 and NGC 507) we chose a higher S/N cutoff to reduce
contamination from the other source.

The background for each galaxy was chosen from source-free regions of
the CCDs.  Although it is preferred to take the background from the
same region of the CCDs as the source in order to insure the same
vignetting correction (i.e. using a background determined from a deep
exposure of a blank field), the variations of the local background due
to environment and peculiarities for a particular observation (e.g.
contamination from the bright Earth, the Sun, the South Atlantic
Anomaly) outweigh the vignetting consideration for the SIS. However,
since the background increases markedly with distance from the center
of the field of the GIS \cite{ishisaki}, we used the blank fields for
the background estimates of the GIS data.  In any event, the results
using a local background generally differ unimportantly from those
when a background template constructed from a deep observation of a
blank field is used (e.g. Buote \& Canizares 1997)\footnote{Typically
the fluxes show the largest differences between the blank sky
templates and a local background. The fitted temperatures and
abundances are mostly unaffected.}.

Finally, the source and background spectra for each galaxy were
extracted in 512 pulse invariant (PI) bins after correcting for
temporal SIS gain variations and the degradation due to
charge-transfer inefficiency.  We mention that we made the required
corrections to fix problems in the response matrices generated by the
standard reduction software as described on the {\sl ASCA} Guest
Observer Facility WWW pages (as of May 1997). Also, the GIS background
templates were rise-time filtered (i.e. with FTOOL {\it gisclean}) to
match the screening of the galaxy observations.

\begin{table*}
\begin{minipage}{133mm}
\caption{{\em ASCA} SIS Observation Properties}
\label{tab.obs}
\begin{tabular}{llrrrrrrrr}
Name  & Sequence \# & \multicolumn{2}{c}{Exposure} &
      \multicolumn{2}{c}{Count Rate} & \multicolumn{2}{c}{$R$} &
      \multicolumn{2}{c}{$R$} \\
      &             &  \multicolumn{2}{c}{($10^{3}$s)}  &
      \multicolumn{2}{c}{($10^{-2}$ ct s$^{-1}$)} &
      \multicolumn{2}{c}{(arcmin)} & \multicolumn{2}{c}{(kpc)}\\
      &             & SIS0 & SIS1 & SIS0 & SIS1 & SIS0 & SIS1 & SIS0 & SIS1\\
NGC 499  & 61007000 & 25.3 & 25.6 & 7.5 & 7.0 & 2.9 & 2.8 & 74.7 & 72.2\\
      & 63026000 & 34.9 & 33.7 & 11.6 & 9.1 & 3.6 & 3.4 & 92.8 & 87.6\\
NGC 507  & 61007000 & 25.3 & $\cdots$ & 22.0 & $\cdots$ & 4.2 &
      $\cdots$ & 121.3 & $\cdots$\\
NGC 720  & 60004000 & 30.2 & 32.0 & 2.5 & 1.4 & 2.5 & 2.0 & 23.7 & 19.0\\
NGC 1332 & 63022000 & 53.6 & 53.1 & 1.8 & 1.0 & 2.7 & 2.1 & 22.3 & 17.3\\
NGC 1399 & 80038000 & 14.9 & 10.3 & 34.4 & 33.8 & 4.9 & 4.4 & 38.9 & 34.8\\ 
      & 80039000 & 14.5 & 13.1 & 36.2 & 13.9 & 4.6 & 3.2 & 36.3 & 25.3\\ 
NGC 1404 & 80038000 & 14.9 & 10.3 & 8.2 & 8.9 & 2.7 & 2.3 & 21.4 & 18.2\\  
      & 80039000 & 14.5 & 13.1 & 8.0 & 4.0 & 2.5 & 2.0 & 19.8 & 15.8\\  
NGC 1407 & 63021000 & 31.7 & 31.3 & 4.7 & 3.7 & 3.0 & 2.7 & 30.3 & 27.3\\
NGC 3923 & 82016000 & 20.5 & 12.7 & 1.7 & 1.3 & 2.0 & 2.0 & 24.1 & 24.1\\ 
NGC 4374 & 60007000 & 13.5 & 12.7 & 4.0 & 4.4 & 2.0 & 2.0 & 15.7 & 15.7\\
NGC 4406 & 60006000 (4ccd) & 11.0 & 9.1 & 38.0 & 26.2 & 5.5 & 5.0 &
      43.2 & 39.3\\ 
      & 60006000 (2ccd) & 5.3  & 5.2 & 36.2 & 20.3 & 5.3 & 4.6 & 41.6
      & 36.1\\
NGC 4472 & 60029000 & 16.3 & 12.5 & 36.2 & 25.7 & 4.1 & 3.5 & 32.2 & 27.5\\ 
      & 60030000 & 17.0 & 13.2 & 32.3 & 20.8 & 4.3 & 2.4 & 33.8 & 18.8\\
NGC 4636 & 60032000 & 33.2 & 24.8 & 40.6 & 27.6 & 5.0 & 4.6 & 39.3 & 36.1 \\
NGC 4649 & 60006000 (2ccd) & 16.7 & $\cdots$ & 13.9 &  $\cdots$ & 3.0
      & $\cdots$ & 23.6 & $\cdots$\\
      & 61005000 (4ccd) & 19.2 & $\cdots$ & 13.8 &  $\cdots$ & 3.0 &
      $\cdots$ & 23.6 & $\cdots$\\
NGC 5044 & 80026000 + 80026010 & 16.3 & 12.2 & 117.4 & 78.4 & 5.5 &
      5.1 & 100.0 & 92.3\\
NGC 5846 & 61012000 & 30.3 & 16.7 & 19.9 & 11.5 & 4.0 & 3.5 & 53.3 & 46.6\\
NGC 6876 & 81020000 & 18.6 & $\cdots$ & 2.5  & $\cdots$ & 2.0 &
      $\cdots$ & 53.4 & $\cdots$\\
NGC 7619 & 63017000 & 52.5 & 48.5 & 4.7 & 3.4 & 3.0 & 3.0 & 54.5 & 54.5\\
NGC 7626 & 63017000 & 52.5 & 48.5 & 0.9 & 1.1 & 2.0 & 2.0 & 36.3 & 36.3\\
IC 1459 & 60005000 & 23.8 & 17.4 & 3.9 & 1.7 & 2.5 & 2.0 & 23.3 & 18.7\\
IC 4296 & 61006000 & 32.9 & 31.3 & 4.7 & 2.7 & 3.1 & 2.6 & 70.7 & 59.3\\
\end{tabular}

\medskip

The exposures include any time filtering. The count rates are
background subtracted within the circle of radius $R$. For NGC 4406 we
give the half-diagonal of the rectangular regions.  Aperture radii in
kpc are computed using the distances from Faber et al. \shortcite{7s}
as reported in Eskridge, Fabbiano, \& Kim \shortcite{efk} for $H_0=50$
km/s/Mpc.
	
\end{minipage}
\end{table*}

\section{Spectral fitting}
\label{fit}

It has been known since the operation of the {\it Einstein
Observatory} that the X-ray emission of early-type galaxies mostly
originates from a hot interstellar medium \cite{f79}, with an
additional contribution from the integrated emission from X-ray
binaries which are most important for the galaxies with the smallest
ratios of X-ray to blue-band optical luminosity \cite{cft}.  The
results from the Performance-Verification phase of {\sl ASCA} showed
that the brightest ellipticals in X-rays have emission lines
consistent with those originating from a thin hot thermal plasma
\cite{awaki}.  The form of the spectrum arising from discrete sources
is not well constrained, but is consistent with a high-temperature
$T\ga 5$ keV thin thermal plasma (Kim et al. 1992; Matsumoto et
al. 1997).

Hence, we concentrate on thin thermal models modified by
photo-electric absorption arising from the ISM of our galaxy. We allow
for the possibility of intrinsic absorption as seen in galaxy clusters
\cite{w91} if the fit to the galaxy spectrum is significantly
improved. Since the hot gas need not be in a single phase and the
X-ray emission may contain a significant contribution from discrete
sources, we fit as many absorbed temperature components as are
required by the {\sl ASCA} data. (As stated in \S \ref{intro}, we do
not assume the presence of a discrete (or other hard) component with a
predetermined functional form for all galaxies.)

To model the emission of a thin thermal plasma we primarily use the
MEKAL code which is a modification of the original MEKA code (Mewe,
Gronenschild, \& van den Oord 1985; Kaastra \& Mewe 1993) where the
Fe-L shell transitions crucial to the X-ray emission of ellipticals
have been re-calculated \cite{mekal}. We take solar abundances
according to Anders \& Grevesse \shortcite{ag} and photo-electric
absorption cross sections according to Baluci\'{n}ska-Church \&
McCammon \shortcite{phabs}. Although not as accurate as the MEKAL
model we also employ the Raymond-Smith code \cite{rs} since it has
been used in most previous studies. 

We also consider a cooling-flow model based on the MEKAL code where
the gas cools continuously from some upper temperature, $T_{max}$. The
emissivity of the cooling gas is described in Johnstone et
al. \shortcite{rjcool}. Although this model may be more appropriate
for the cores of galaxy clusters since it does not include potential
heat sources which may be important in galaxies, it provides a simple
and convenient exploration of the effects of cooling-flow spectra on
the {\sl ASCA} galaxy data. The cooling flow spectrum assumes that the
gas cools at constant pressure. The pressure itself does not enter
into the calculation, and the same spectrum results from different
parcels of gas cooling at different pressures. Provided that the gas
at different radii (and thus different pressures) starts from the same
temperature and cools before flowing a significant distance, it is
therefore a good approximation. In practice, gravitational work will
be done as the gas flows, so the true result will be more complicated
(in a manner which depends on the degree of flow). An additional
temperature component is the simplist modification that can be made
for this effect (see \S 3.2.3).

All of the spectral fitting was implemented with the software package
XSPEC \cite{xspec} using the $\chi^2$ minimization method. In order
for the weights to be valid for the $\chi^2$ method we regrouped the
PI bins such that each group contained at least 20 counts. We
restricted the fits to energies above 0.5 keV ($\sim 0.55$ keV)
because of uncertainties in the SIS response at lower energies.  We
also exclude energies $>5$ keV because the $S/N$ of the bins at these
energies is generally too low to significantly affect the fits and
these energies tend to be dominated by uncertainties in the background
level of the SIS and to calibration errors \cite{gendreau}. The
normalizations of the SIS0 and SIS1 are fitted separately as are the
normalization of multiple observations of the same galaxy. (For the
GIS data we analyzed energies $0.8-9$ keV.)

For NGC 507 we found that the spectral shapes of the SIS0 and SIS1
differed substantially. However, we find no such discrepancy for the
NGC 499 data taken from the NGC 507 observation. (The N499 and N507
data are take from different chips of the SIS.)  We compared the NGC
507 data with those from a lower S/N observation (61007010) and found
that the SIS0 spectra are consistent. (The SIS1 data for 61007010 have
too low S/N to make a useful comparison.)  The GIS data give fitted
temperatures and abundances and reduced $\chi^2$ in excellent
agreement with the SIS0 data. Thus we restrict our analysis of NGC 507
to the SIS0 data. (We mention that we find no similar SIS0/SIS1
discrepancy for the remaining 19 galaxies.)

\subsection{Single-component fits}
\label{1t}

\subsubsection{Results}

\renewcommand{\arraystretch}{1.35}

\begin{table*}
\begin{minipage}{155mm}
\caption{Single-Component Fits}
\label{tab.1t}
\begin{tabular}{llccccrrrr}
Name & Model & $N_{\rm H}$ & $T$ & $Z$ & $\dot{M}_{\rm gas}$ &
$\Delta\chi^2(N_{\rm H})$ & $\chi^2$ & dof & $\chi^2_{\rm red}$\\
     &       & ($10^{21}$ cm$^{-2}$) & (keV) & $(Z_{\sun})$ &
$M_{\sun}$ yr$^{-1}$\\ 
NGC 499  & MEKAL & 2.0 & 0.67 & 0.31 &      & 49.4 & 323.3 & 205 & 1.58\\
      & CF    & 2.4 & 0.96 & 0.48 & 45.9 & 122.0 & 334.6 & 205 & 1.63\\
NGC 507  & MEKAL & $1.2_{-0.4}^{+0.3}$ & $1.27_{-0.06}^{+0.06}$ &
$0.38_{-0.08}^{+0.10}$ & & 12.3 & 68.4 & 74 & 0.92\\
      & CF    & $3.0_{-0.6}^{+0.5}$ & $2.16_{-0.20}^{+0.27}$ &
$1.9(>1.3)$ & $34.5_{-7.7}^{+8.3}$ & 96.1 & 80.2 & 74 & 1.08\\
NGC 720  & MEKAL & $\cdots$ & 0.63 & 0.08 & & $\cdots$ & 94.2 & 48 & 1.96\\
      & CF    & $\cdots$ & 1.11 & 0.12 & 0.94 &$\cdots$ & 90.9 & 48 & 1.89\\
NGC 1332 & MEKAL & $\cdots$ & 0.65 & 0.06 & & $\cdots$ & 173.6 & 66 & 2.63\\
      & CF    & $\cdots$ & 1.17 & 0.08 & 0.49 &$\cdots$  & 169.1 & 66 & 2.56\\
NGC 1399 & MEKAL & 0.6 & 1.27 & 0.37 &  & 29.9 & 435.8 & 256 & 1.70\\
      & CF    & $1.8_{-0.2}^{+0.3}$ & $2.5_{-0.1}^{+0.2}$ &  $(>1.7)$
& $1.14_{-0.12}^{+0.13}$ & 233.7 & 341.3 & 256 & 1.33\\
NGC 1404 & MEKAL & 1.4 & 0.56 & 0.27 & & 20.6 & 196.6 & 127 & 1.55\\
      & CF    & 2.1 & 0.73 & 0.44 & 13.9  & 39.6 & 197.5 & 127 & 1.55\\
NGC 1407 & MEKAL & $\cdots$ & 0.92 & 0.10 & &$\cdots$ & 166.4 & 94 & 1.77\\
      & CF     & $\cdots$ & $2.4_{-0.2}^{+0.3}$ &
$0.38_{-0.11}^{+0.14}$ & $0.63_{-0.10}^{+0.09}$ &$\cdots$  & 138.4 & 94 &
1.47\\ 
NGC 3923 & MEKAL & $\cdots$ & 0.64 & 0.07 &  &$\cdots$  & 48.3 & 23 & 2.10\\
      & CF    & $\cdots$ & 1.13 & 0.11 & 0.84  &$\cdots$  & 45.3 & 23 & 1.97\\
NGC 4374 & MEKAL & $\cdots$ & $0.70_{-0.04}^{+0.05}$ &
$0.09_{-0.03}^{+0.04}$ &  &$\cdots$  & 72.2 & 52 & 1.39\\
      & CF    & $\cdots$ & $1.17_{-0.13}^{+0.16}$ &
$0.13_{-0.04}^{+0.08}$ & $0.55_{-0.13}^{+0.15}$ &$\cdots$ &  74.7 & 52 & 1.44\\
NGC 4406 & MEKAL & $1.2_{+0.4}^{-0.3}$ & $0.73_{-0.02}^{+0.03}$ &
$0.27_{-0.04}^{+0.06}$ &  & 23.4 & 223.5 & 188 & 1.19\\
      & CF    & $2.3_{-0.3}^{+0.4}$ & $1.03_{-0.06}^{+0.07}$ &
$0.69_{-0.20}^{+0.34}$ & $4.7_{-0.8}^{+1.0}$ & 116.1 & 213.6 & 188 & 1.14\\
NGC 4472 & MEKAL & 0.8 & 0.97 & 0.26 &  & 53.6 & 495.4 & 240 & 2.06\\
      & CF    & $2.2_{-0.3}^{+0.1}$ & $1.59_{-0.06}^{+0.06}$ &
$1.31_{-0.23}^{+0.31}$ & $1.94_{-0.18}^{+0.19}$ & 338.5 & 301.1 & 240
& 1.26\\
NGC 4636 & MEKAL & 1.2 & 0.66 & 0.28 &  & 86.9 & 280.6 & 138 & 2.03\\  
      & CF    & 1.9 & 0.91 & 0.48 & 5.61 & 245.7 & 295.3 & 138 & 2.14\\
NGC 4649 & MEKAL & 1.6 & 0.83 & 0.20 &  & 51.3 & 210.1 & 99 & 2.12\\
      & CF    & 2.3 & 1.39 & 0.56 & 1.37 & 99.8 & 179.7 & 99 & 1.82\\
NGC 5044 & MEKAL & 0.9 & 0.95 & 0.25 &   &  30.8 & 443.9 & 161 & 2.76\\
      & CF    & $2.5_{-0.2}^{+0.2}$ & $1.42_{-0.04}^{+0.05}$ &
$1.03_{-0.15}^{+0.17}$ & $75.7_{-6.5}^{+6.8}$ & 542.5 & 228.5 & 161 &
1.42\\
NGC 5846 & MEKAL & $1.7_{-0.4}^{+0.6}$ & $0.67_{-0.03}^{+0.04}$ &
$0.21_{-0.04}^{+0.06}$ & & 29.5 & 139.9 & 107 & 1.31\\
      & CF    & $2.0_{-0.4}^{+0.4}$ & $1.02_{-0.06}^{+0.10}$ &
$0.38_{-0.08}^{+0.09}$ & $9.4_{-2.3}^{+3.0}$ & 51.1 & 109.4 & 107 &
1.02\\ 
NGC 6876 & MEKAL & $2_{-1}^{+4}$ & $0.88_{-0.25}^{+0.15}$ &
$0.14_{-0.07}^{+0.57}$ & & 3.3 & 22.6 & 17 & 1.33\\
      & CF    & $3.1_{-1.7}^{+2.8}$ & $1.3_{-0.4}^{+0.6}$ &
$0.3_{-0.2}^{+1.1}$ & $8.8_{-5.5}^{+21}$ & 8.1 & 20.5 & 17 & 1.20\\
NGC 7619 & MEKAL & $3.1_{-0.9}^{+0.9}$ & $0.69_{-0.5}^{+0.5}$ &
$0.31_{-0.10}^{+0.23}$ &  & 44.4 & 117.3 & 98 & 1.20\\
      & CF    & $3.3_{-0.8}^{+0.5}$ & $1.03_{-0.10}^{+0.14}$ &
$0.46_{-0.13}^{+0.25}$ & $17.5_{-5.8}^{+6.2}$ & 70.4 & 112.1 & 98 &
1.14\\ 
NGC 7626 & MEKAL & $\cdots$ & 0.88 & 0.08 &  &$\cdots$  & 85.6 & 52 & 1.65\\
      & CF    & $\cdots$ & $2.31_{-0.44}^{+0.54}$ &
$0.38_{-0.19}^{+0.39}$ & $0.72_{-0.19}^{+0.26}$ &$\cdots$ & 70.2 &  52 & 1.35\\
IC 1459 & MEKAL & $\cdots$ & 2.69 & 0.00 &  &$\cdots$  & 100.3 & 55 & 1.82\\
      & CF    & $\cdots$ & $7.4_{-1.7}^{+3.0}$ &
$0.11_{-0.06}^{+0.13}$ & $0.14_{-0.04}^{+0.04}$ &$\cdots$  & 83.5 & 55 &
1.52\\
IC 4296 & MEKAL & $\cdots$ & 1.02 & 0.08 &   &$\cdots$   & 203.2 & 91 & 2.23\\
      & CF    & $\cdots$ & $3.3_{-0.5}^{+0.7}$ &
$0.47_{-0.16}^{+0.20}$ & $1.83_{-0.36}^{+0.37}$ & $\cdots$ & 137.0 & 91 & 1.49\\
\end{tabular}
\medskip

Results of fitting either one MEKAL model or a cooling-flow model each
modified by photo-electric absorption. If allowing $N_{\rm H}$ to be
free significantly improved the fit, we give the best-fit value of
$N_{\rm H}$ and the change in $\chi^2$ denoted by $\Delta\chi^2(N_{\rm
H})$. Otherwise ``$\cdots$'' denote that $N_{\rm H}$ is kept at its
Galactic value.  For fits where $\chi^2<1.5$ we also list 90\%
confidence limits on one interesting parameter. The gas mass
deposition rates, $\dot{M}_{\rm gas}$, are quoted in the 0.5-5 keV
band for the SIS0. For galaxies with multiple observations, the
highest S/N observation is typically quoted.

\end{minipage}
\end{table*}

\renewcommand{\arraystretch}{1.0}

\begin{figure*}
\parbox{0.49\textwidth}{
\centerline{\psfig{figure=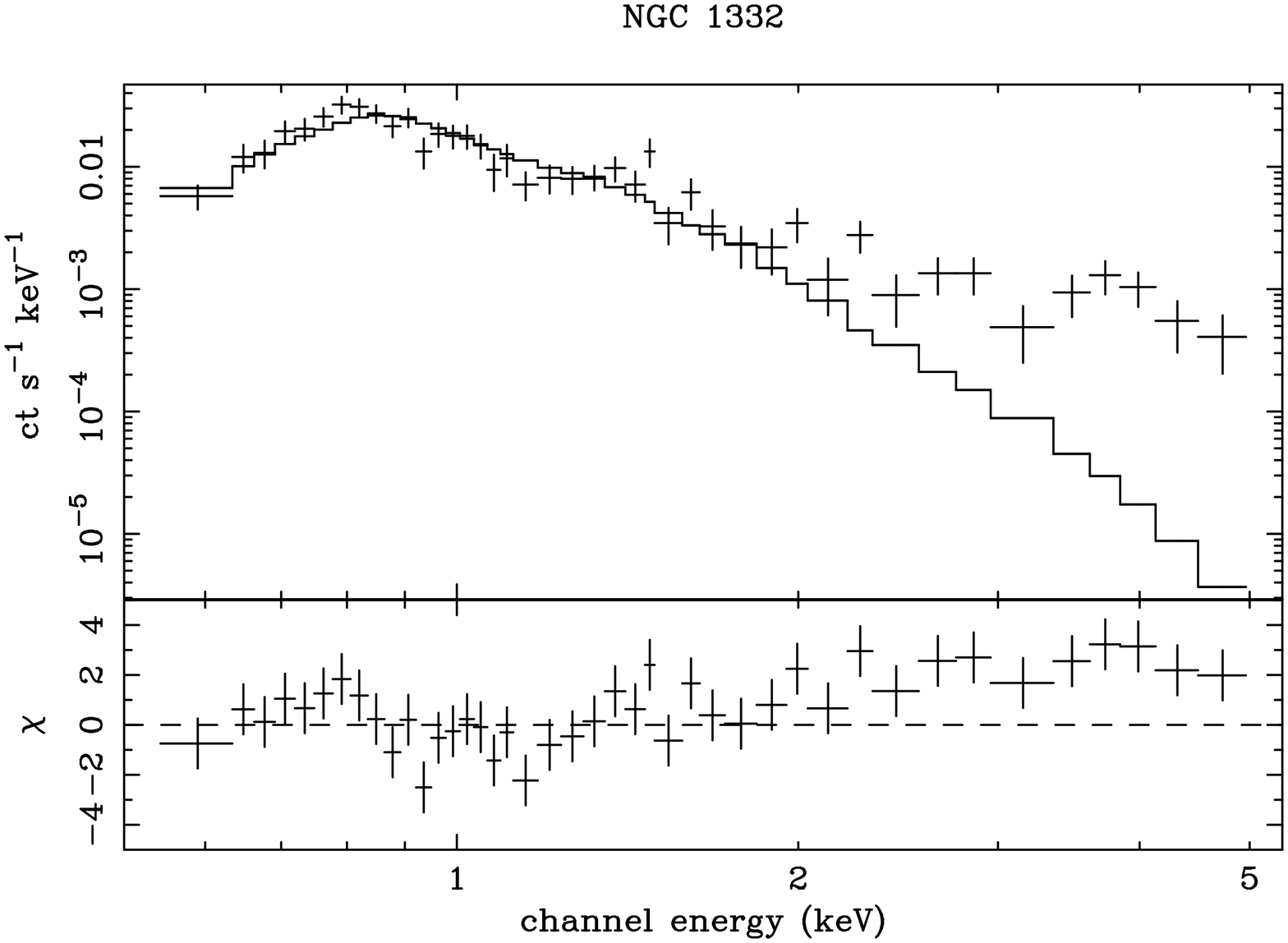,angle=-90,height=0.3\textheight}}
}
\parbox{0.49\textwidth}{
\centerline{\psfig{figure=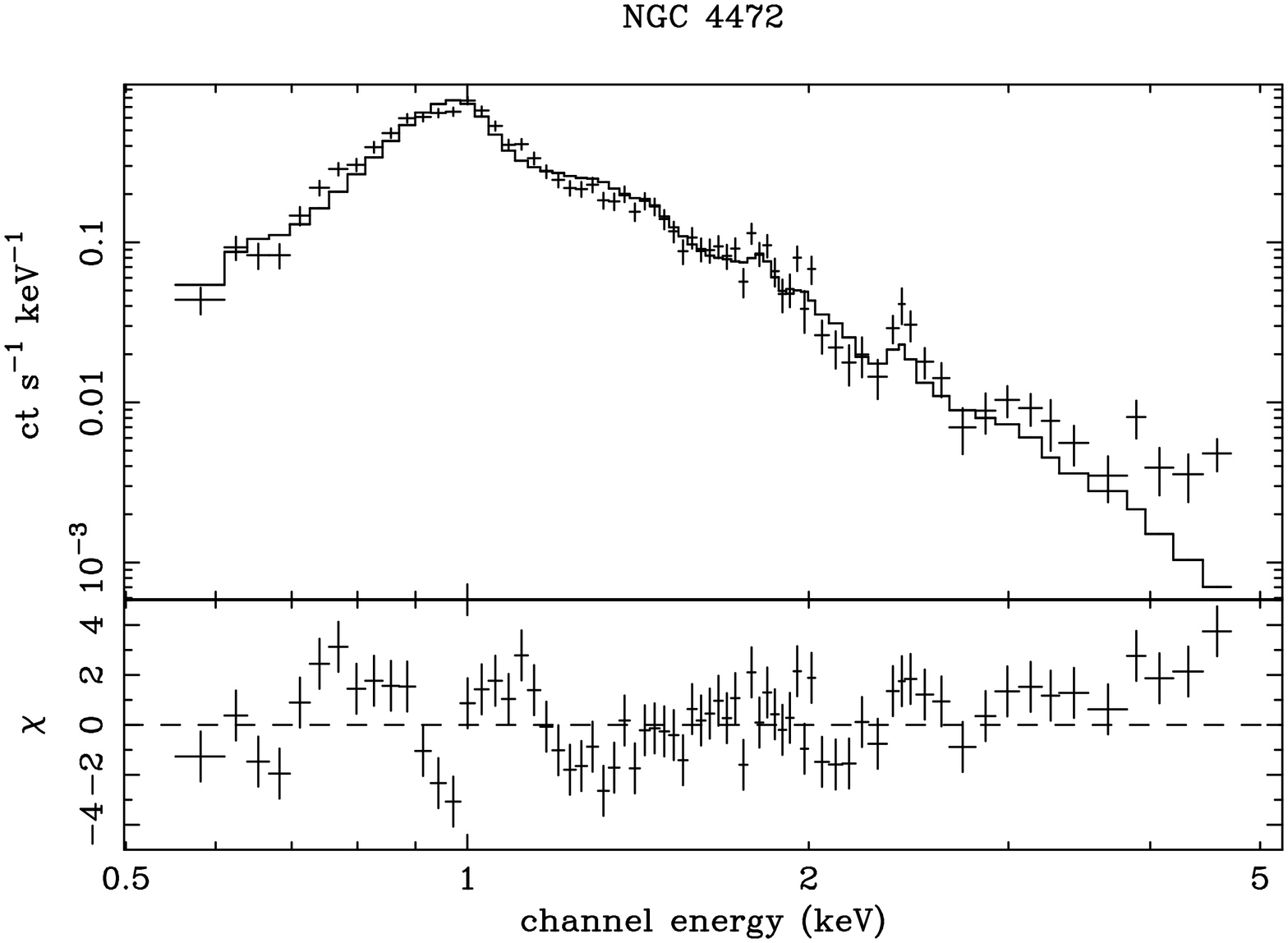,angle=-90,height=0.3\textheight}}
}
\vspace{-0.2cm}
\parbox{0.49\textwidth}{
\centerline{\psfig{figure=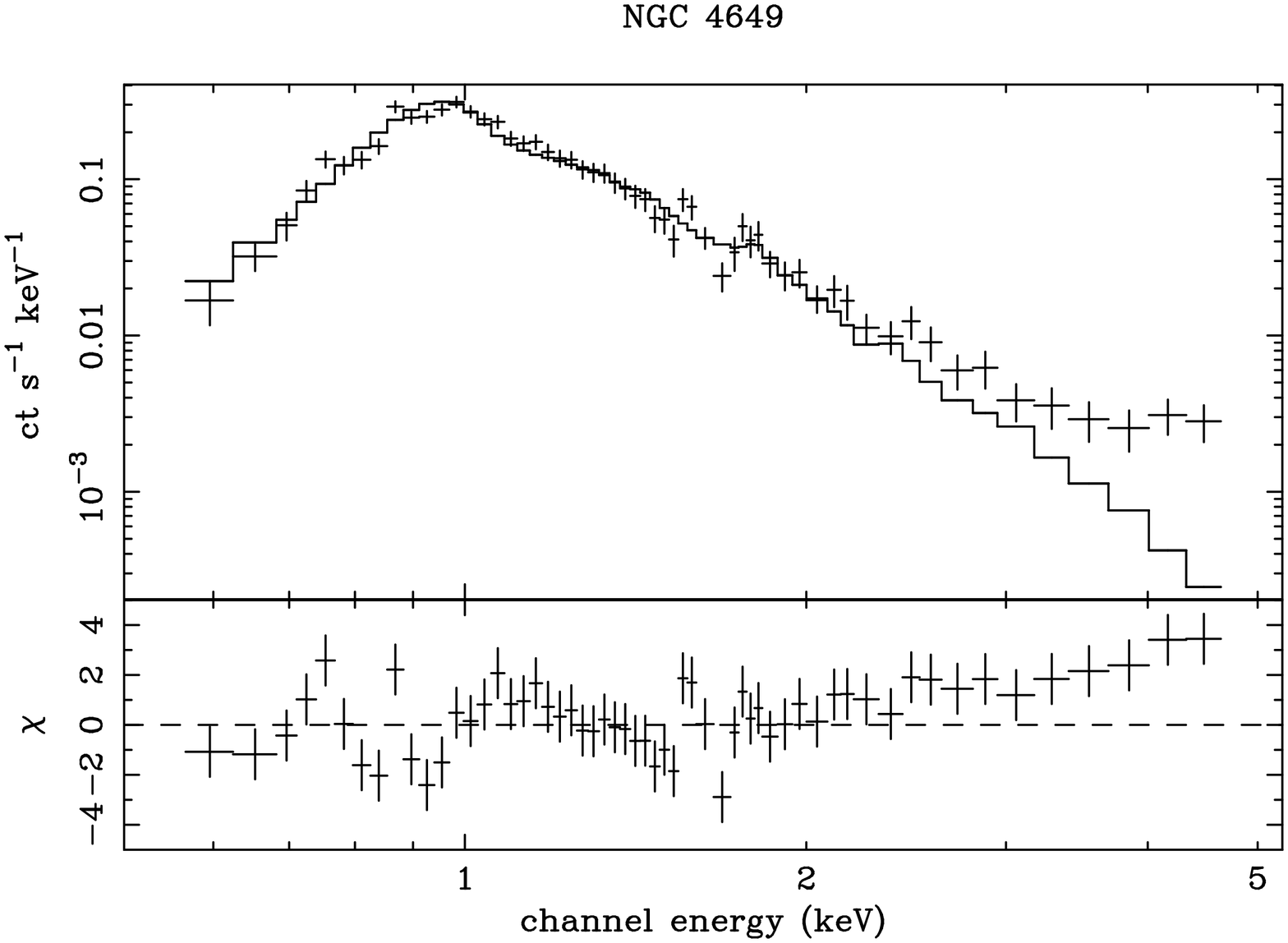,angle=-90,height=0.3\textheight}}
}
\parbox{0.49\textwidth}{
\centerline{\psfig{figure=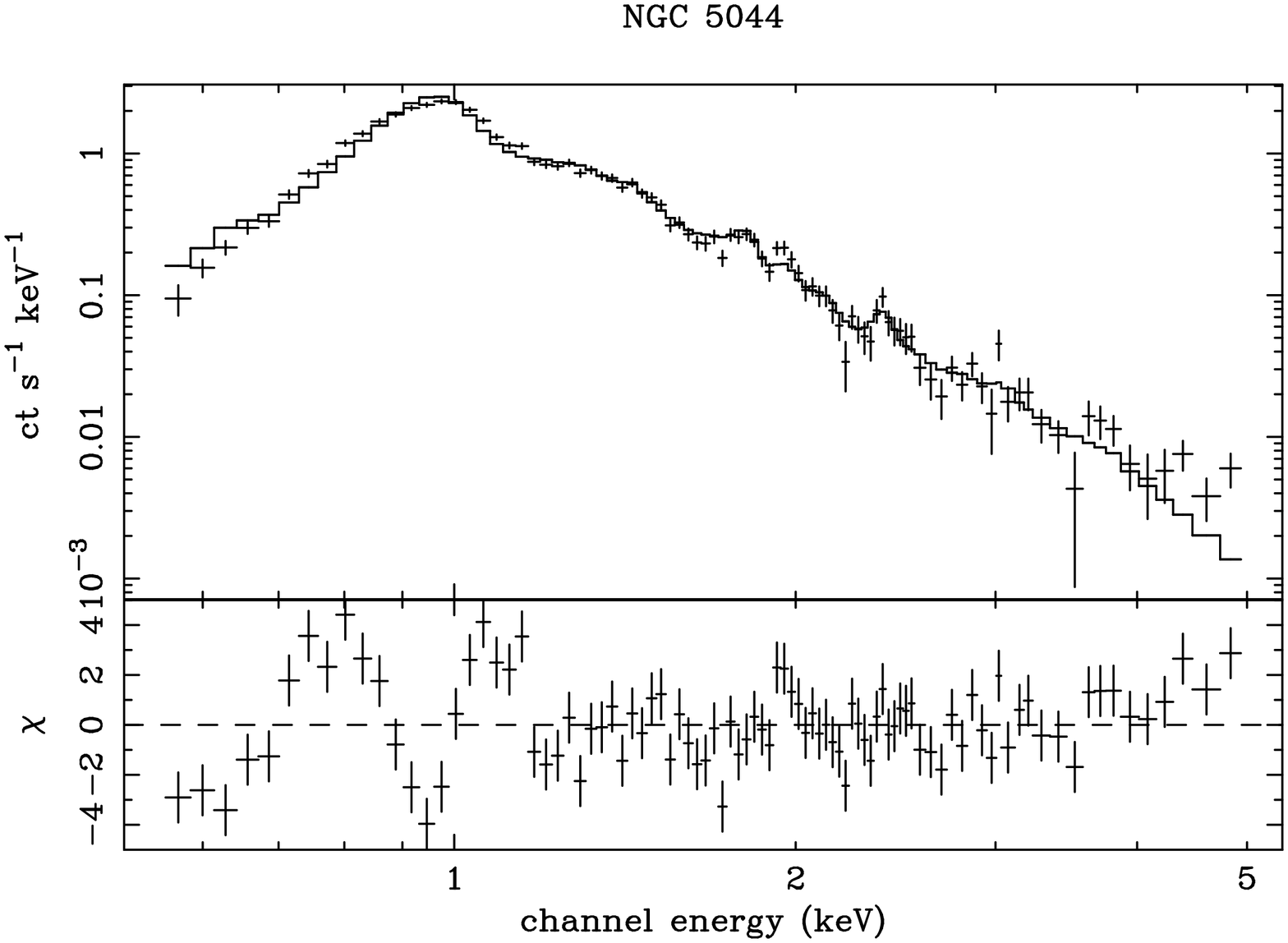,angle=-90,height=0.3\textheight}}
}
\caption{\label{fig.1t} Absorbed single-temperature (MEKAL) fits to
the spectra of NGC 1332, NGC 4472, NGC 5044, and NGC 4649. Only the
SIS0 data is shown.}
\end{figure*}

In Table \ref{tab.1t} we list the results of simultaneously fitting a
single absorbed MEKAL model to the available SIS0 and SIS1 spectra of
each galaxy listed in Table \ref{tab.obs}.  In Figure \ref{fig.1t} we
show the spectral fits of NGC 1332 and NGC 4649 (among best S/N for
lowest $L_{\rm x}/L_{\rm B}$) and NGC 4472 and NGC 5044 (among best
S/N for highest $L_{\rm x}/L_{\rm B}$).  (For those galaxies with
multiple observations we show the observation with the longest
exposure.)  In almost all cases the fits are formally quite poor with
reduced $\chi^2 > 1.5$. For the X-ray brightest galaxies (and those
with largest $L_{\rm x}/L_{\rm B}$) the residuals of the fits tend to
be most pronounced around 1~keV, whereas the fainter galaxies (and
those with smallest $L_{\rm x}/L_{\rm B}$) tend to show the largest
significant residuals at higher energies, $E\ga 3$ keV.

The temperatures and abundances determined from the single-temperature
fits are qualitatively consistent with previous {\sl ASCA} studies of
the ellipticals in our sample (Awaki et al. 1994; Matsushita et
al. 1994; Loewenstein et al. 1994; Fukazawa et al. 1996; Buote \&
Canizares 1997; Arimoto et al. 1997; Matsumoto et al. 1997; Iwasawa,
White, \& Fabian 1997) modulo expected small systematic differences
due to different source region sizes, background region definitions,
thermal emission codes used, current status of SIS calibration etc. In
particular, our single-temperature models reproduce the extremely
sub-solar abundances found in the above studies. All the galaxies have
best-fit $Z<0.4Z_{\sun}$ and most have $Z\la 0.15Z_{\sun}$. The mean
abundance and standard deviation are $\langle Z\rangle =0.19\pm 0.12$
$Z_{\sun}$.

For half of the galaxies in our sample the fits improved significantly
if $N_{\rm H}$ was allowed to be a free parameter. This improvement
affects only the energy bins for $E\sim 0.5-1.0$ keV.  In each of
these cases, significant excess absorption above the Galactic value is
indicated.  The excess absorption we find for these galaxies agrees
qualitatively with those determined from the previous {\sl ASCA}
studies listed above.  However, with {\sl ROSAT} only NGC 1399 showed
evidence for excess absorption \cite{vijay} while several of the
galaxies which we find to require excess absorption were not found to
require it with {\it ROSAT}: NGC 499 and NGC 507 \cite{kf}, NGC 4472
(Forman et al. 1993; Irwin \& Sarazin 1996), NGC 4636 \cite{trin}, NGC
4649 and NGC 7619 \cite{tfk}, NGC 5044 \cite{david}, and several
others \cite{dw}. We discuss in detail the evidence and implications
for this excess absorption in \S \ref{nh}.

In terms of both $\chi^2$ and the fitted parameters the Raymond-Smith
model gives broadly comparable results to the MEKAL model when fitted
over the $\sim 0.5-5$ keV range.  (In agreement with many previous
studies we find that the Raymond-Smith model tends to give higher
temperatures and smaller abundances than the MEKAL model.)  As
expected, the models differ in the character of their residuals over
the Fe-L region $\sim 0.7-1.2$ keV. One effect of these differences is
that the Raymond-Smith fits are not usually improved as much by
letting $N_{\rm H}$ be a free parameter.

\begin{figure*}
\parbox{0.49\textwidth}{
\centerline{\psfig{figure=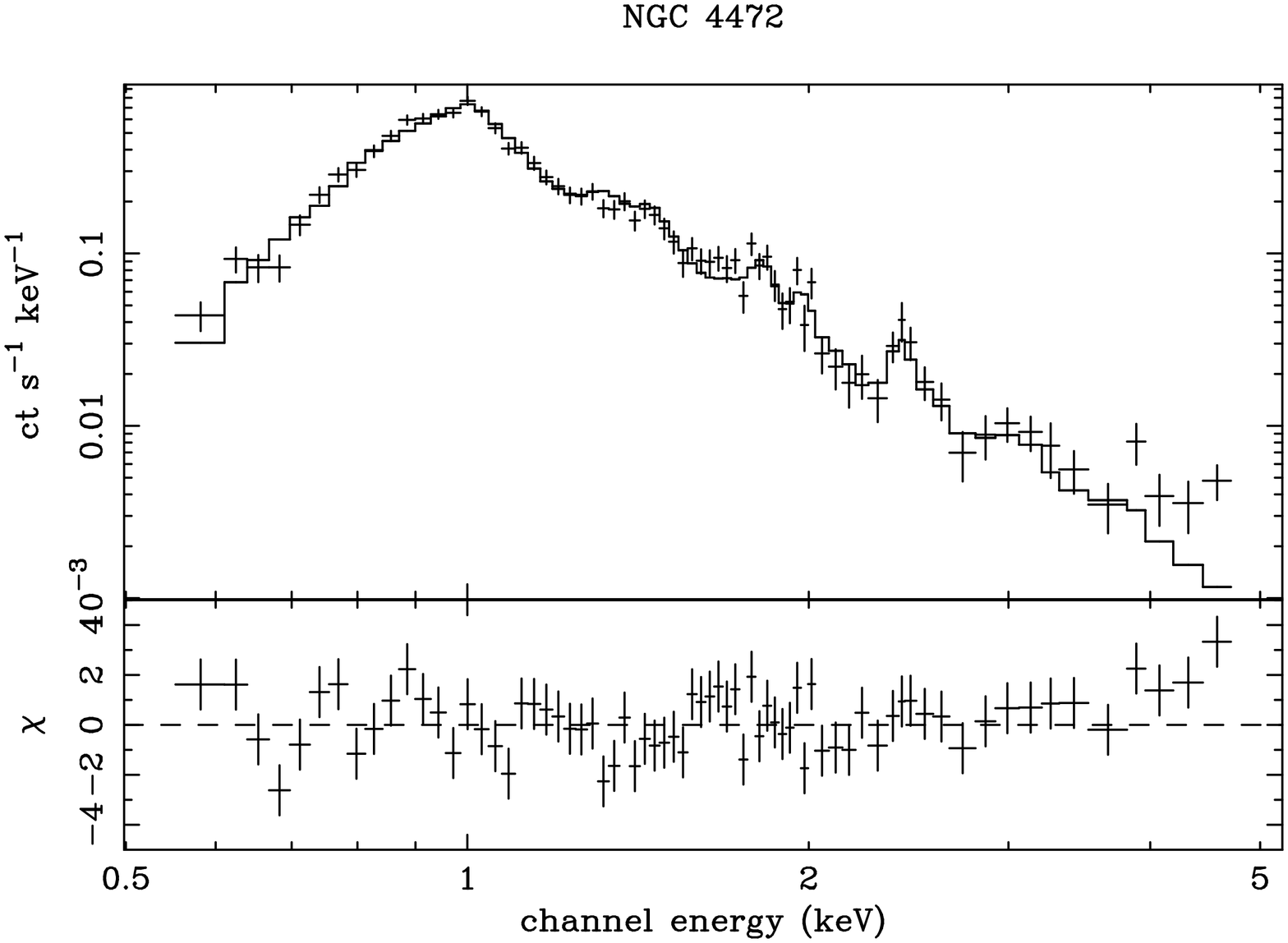,angle=-90,height=0.3\textheight}}
}
\parbox{0.49\textwidth}{
\centerline{\psfig{figure=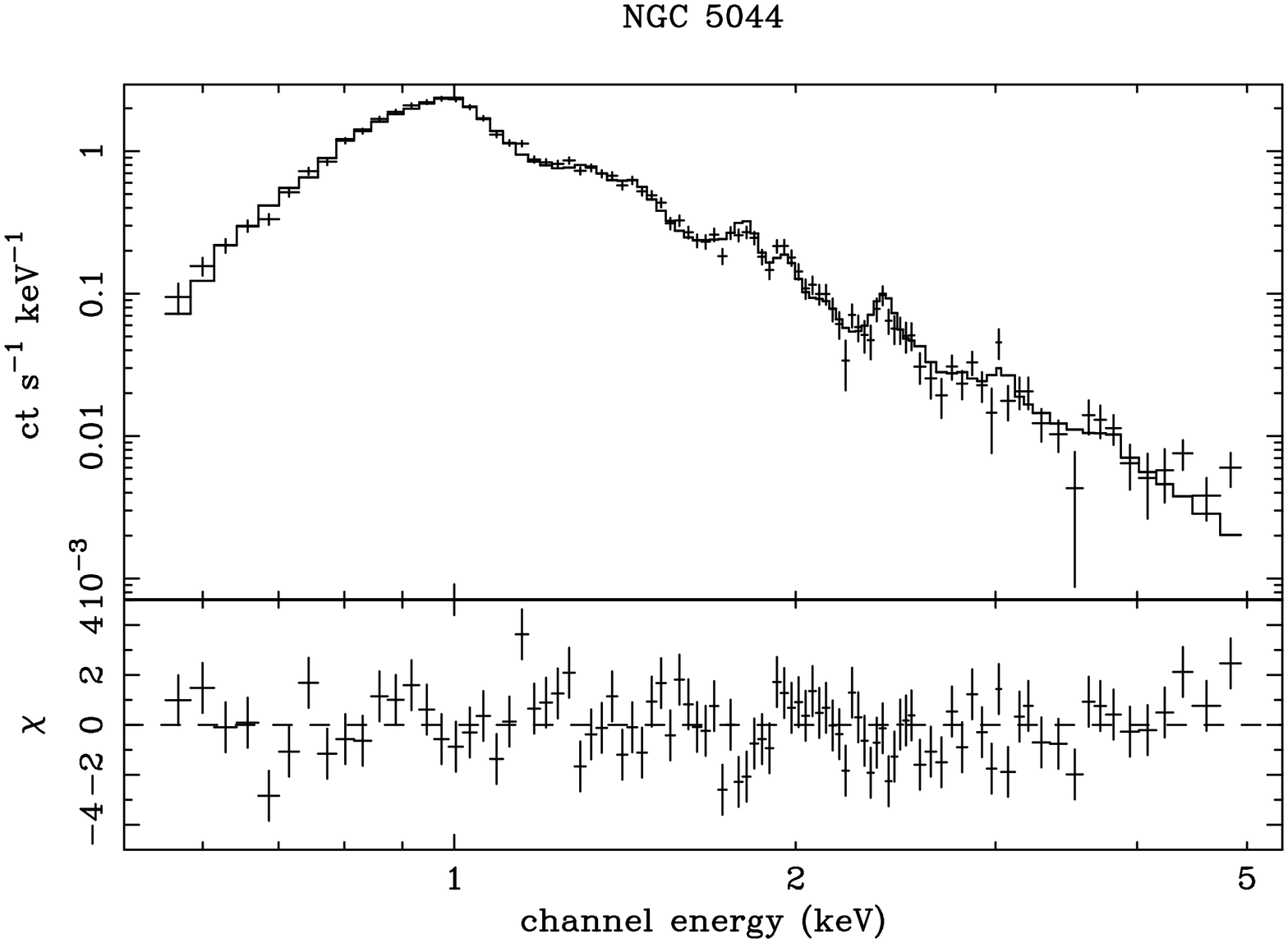,angle=-90,height=0.3\textheight}}
}
\caption{\label{fig.cf} Fits of the absorbed cooling-flow model to
NGC 4472 (left) and NGC 5044 (right). Only the SIS0 data is shown. }
\end{figure*}

We also display in Table \ref{tab.1t} the results for the absorbed
cooling-flow model.  For eleven galaxies the cooling-flow model with
or without excess absorption fits the spectra with $\chi^2$ values
comparable to the MEKAL model.  For the other nine galaxies (IC 1459,
IC 4296, NGC 1399, NGC 1407, NGC 4472, NGC 4649, NGC 5044, NGC 5846,
NGC 7626) the cooling-flow model fits the galaxy spectra better than
the MEKAL model. In the cases of IC 4296, NGC 4472, and NGC 5044 the
improvement in the fits is dramatic (see Figure \ref{fig.cf}).  For
the galaxies where allowing for excess absorption improved the fits to
the MEKAL model, the fits with the cooling-flow model also improved
with excess absorption and typically required a larger amount of
absorption than did the MEKAL model.

The mass deposition rates of the cooling gas, $\dot{M}_{\rm gas}$, are
mostly a few solar masses per year consistent with results from {\it
Einstein} (Nulsen, Stewart, \& Fabian 1984; Thomas et al. 1986).  We
find generally good agreement with the mass deposition rates obtained
from {\sl ROSAT} studies for NGC 507 \cite{kf} and NGC 1399
\cite{vijay}, although the Rangarajan et al. value of $\sim
1.7M_{\sun}$ yr$^{-1}$ for NGC 1399 is about 50 per cent larger than
our value. We find a larger value of $\dot{M}_{\rm gas}$ for NGC 4406
than Iwasawa et al. \shortcite{kazu} because our value refers to a
larger extraction region.

For NGC 5044 we obtain a very large $\dot{M}_{\rm gas}\approx
75M_{\sun}$ yr$^{-1}$ more appropriate for clusters of galaxies (e.g.
Fabian 1994). A previous {\sl ROSAT} study \cite{david} of NGC 5044
found $\dot{M}_{\rm gas}\approx 20-25$ $M_{\sun}$ yr$^{-1}$.  The
larger value we obtain for NGC 5044 occurs because of the large amount
of excess absorption required by {\sl ASCA} whereas for {\sl ROSAT}
the absorption is consistent with the Galactic value.

Interestingly, the abundances obtained from the cooling-flow models
are generally much larger than found from the regular MEKAL
models. For the cooling-flow galaxies mentioned above the
metallicities all have best-fit values $Z\ga 0.4Z_{\sun}$ except for
IC 1459. If we consider all of the galaxies, the cooling-flow models
give a mean and standard deviation, $\langle Z\rangle =0.6\pm 0.5$
$Z_{\sun}$.

\begin{figure*}
\parbox{0.49\textwidth}{
\centerline{\psfig{figure=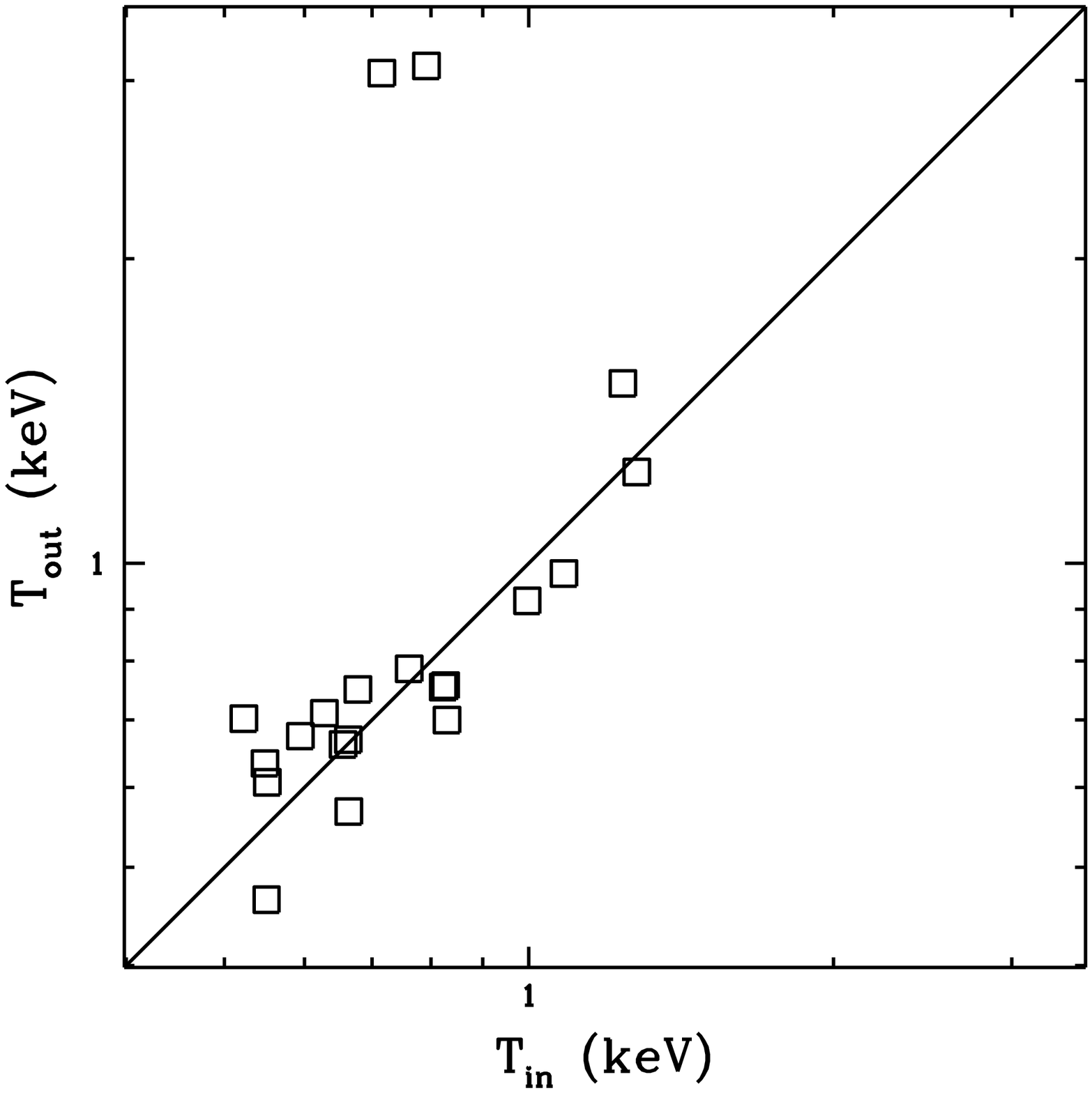,angle=0,height=0.3\textheight}}
}
\parbox{0.49\textwidth}{
\centerline{\psfig{figure=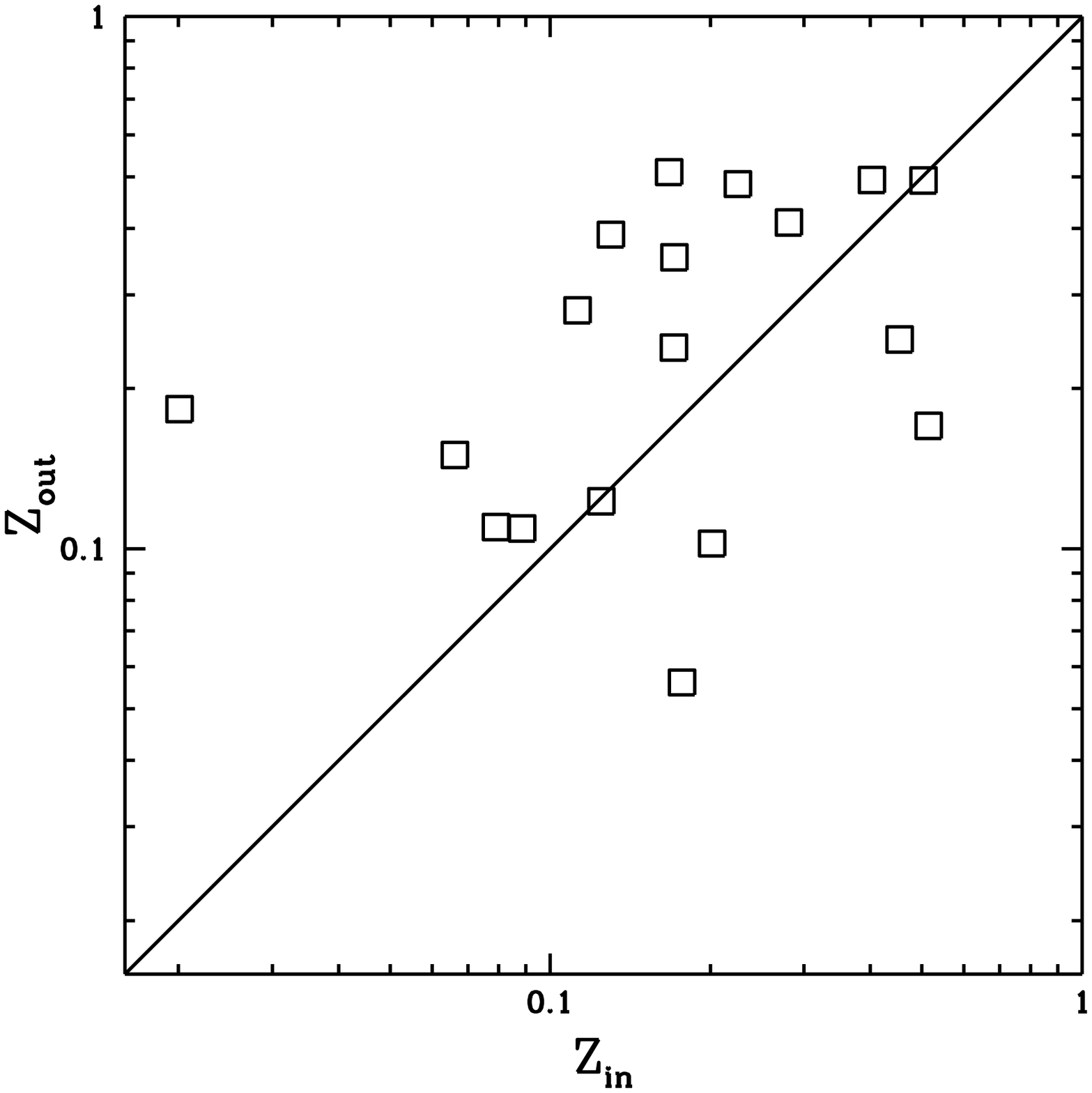,angle=0,height=0.3\textheight}}
}
\caption{\label{fig.test} Empirical test of the reliability of the
MEKAL code within the Fe-L region.  We show results of fitting an
absorbed single-temperature MEKAL model to subdivisions of the energy
range 0.5-2 keV. We designate the Fe-L region (0.7-1.2 keV) as ``in''
while the region over 0.5-2 keV excluding the Fe-L region is
designated as ``out''. The solid lines denote $T_{\rm in}=T_{\rm out}$ and
$Z_{\rm in}=Z_{\rm out}$ respectively.}
\end{figure*}

We mention that the larger values of $T_{\rm CF}$ with respect to
those derived from the MEKAL model do not necessarily imply different
gravitating masses. The value of $T_{\rm CF}$ is just the upper
temperature from which the gas cools at constant pressure. The proper
temperature to use for determining a virial mass is a (gas)
mass-weighted temperature. It should also be emphasized that there
could exist complicated situations in a multi-phase cooling flow where
cold gas which has dropped out of the flow is supported by the hotter
phases (e.g. via magnetic fields).  Again, a (gas) mass-weighted
temperature, or an appropriately adjusted mean molecular weight (which
will be higher than the single-phase hot gas), must be used in the
hydrostatic equation. This leads to a smaller gravitating mass than
that inferred from the equation of hydrostatic equilibrium using
$T_{\rm CF}$. These comments also apply to the two-temperature models
discussed later (\S \ref{2t}).

Interpretation of the fitted parameters for IC 1459 must be treated
with caution. The high temperature, especially for the cooling-flow
model, implies an implausibly large mass for an elliptical of its
luminosity. Rather, emission from discrete sources with a likely
contribution from a central AGN (due to it being a radio source)
suggest that emission from hot gas does not dominate the
emission. This picture is consistent with the relatively small value
of $L_{\rm x}/L_{\rm B}$ of IC 1459 and the results of two-temperature
fits discussed in \S \ref{2t}.

Finally, when jointly analyzing the SIS+GIS data we obtained
essentially the same answers as when analyzing only the SIS
data. However, some differences when analyzing the GIS data alone
deserve mention. First, for the high $L_{\rm x}/L_{\rm B}$ galaxies
such as NGC 507 and NGC 4472, the required excess absorption depended
sensitively on the lower energy bound used in the fits. For $\rm
E_{low} = 0.8$ keV we obtained best-fit $N_{\rm H}=0.01\times 10^{21}$
cm$^{-2}$ for NGC 507 and $N_{\rm H}=2.3\times 10^{21}$ cm$^{-2}$ for
NGC 4472 whereas for $\rm E_{low} = 1$ keV we obtained $N_{\rm
H}=1.8\times 10^{21}$ cm$^{-2}$ for NGC 507 and $N_{\rm H}=1.6\times
10^{21}$ cm$^{-2}$ for NGC 4472; i.e. better agreement with the SIS
data is seen for $\rm E_{low} = 1$ keV.  Second, for the low $L_{\rm
x}/L_{\rm B}$ galaxies such as NGC 1332 and NGC 3923 we find that the
best-fitted temperatures are approximately 2 keV as opposed to about
0.6 keV with the SIS. These values are formally inconsistent within
their errors and simply reflect the different energy resolutions and
different effective areas as a function of energy of the SIS and GIS
as discussed in \S \ref{obs}.

\subsubsection{Assessing the validity of measured abundances }
\label{abun.1t}

The very sub-solar abundances obtained from the single-temperature
MEKAL and Raymond-Smith fits has cast doubt on the reliability of the
plasma codes, in particular the accuracy to which the emission in the
Fe-L spectral region, $0.7-1.4$ keV, is modeled (e.g. Arimoto et
al. 1997). We have empirically tested the reliability of the MEKAL
code by examining the behavior of the fitted abundances and
temperatures when the Fe-L spectral region is excluded from the fits
to the SIS data.  This procedure is motivated by the fact that for gas
with $T\sim 1$ keV the peak in the galaxy spectrum occurs in the Fe-L
region: the location of this peak is one measure of the
temperature. Outside the Fe-L region the continuum level provides an
independent measure of the temperature.  Comparing the temperatures
(and abundances) determined inside and outside the Fe-L region
directly tests the consistency of the plasma code.

We analyzed the restricted energy range, $E=0.5-2$ keV, to limit the
effects of a possible hotter component in the hot gas and/or a hard
component due to discrete sources. We fitted the MEKAL model with free
$N_{\rm H}$ to the following two cases: (in) -- the Fe-L region and (out) --
the energies excluding the Fe-L region over 0.5-2 keV. For both (in)
and (out) the initial parameters in the fits were specified by the
model obtained from fitting the entire 0.5-2 keV range. For these fits
we defined the Fe-L region to be $E=0.7-1.2$ keV.  Although Fe-L
contributes up to energies $\sim 1.4$ keV, the emission there is
comparable to those from other elements (most notably Mg).

Let us suppose that the MEKAL model produces too much Fe-L emission
for a given $T$ and $Z$\footnote{Relative abundances are fixed at
their solar values.}. The fits can suppress the excess Fe-L emission
by either increasing $T$ or decreasing $Z$ or both. Hence, in this
case we would expect fits to region (in) to have higher $T$ and/or
smaller $Z$ than fits to (out).  We would expect the opposite behavior
if the MEKAL model produces too little Fe-L emission for a given $T$
and $Z$. (These arguments assume a single temperature component
dominates the emission over 0.5-2 keV.)

In Figure \ref{fig.test} we plot temperatures and abundances for
region (out) against those of region (in). For all but two galaxies
(IC 1459, IC 4296) the temperatures show little scatter along the line
$T_{\rm in}=T_{\rm out}$ giving a strong indication of the validity of the
MEKAL code in the Fe-L region. The high inferred temperature of IC 1459
(see Table \ref{tab.1t}) accounts for its large value of $T_{\rm out}$
while the significant contribution from a higher temperature component
over region (out) likely explains the position of IC 4296 (see \S
\ref{2t}).

The abundances show considerable scatter, but there appears to be an
excess of galaxies above the line $Z_{\rm in}=Z_{\rm out}$; i.e. $Z_{\rm out}$
tends to exceed $Z_{\rm in}$ in our sample, though the effect is not
large. Given the consistency of the temperatures, this excess
$Z_{\rm out}$ may indicate that the abundances of the elements dominating
the line emission of region (out) (i.e. Si) are slightly over-abundant
with respect to Fe. Once again IC 1459 is the outlier (leftmost point).

\begin{figure}
\centerline{\hspace{0cm}\psfig{figure=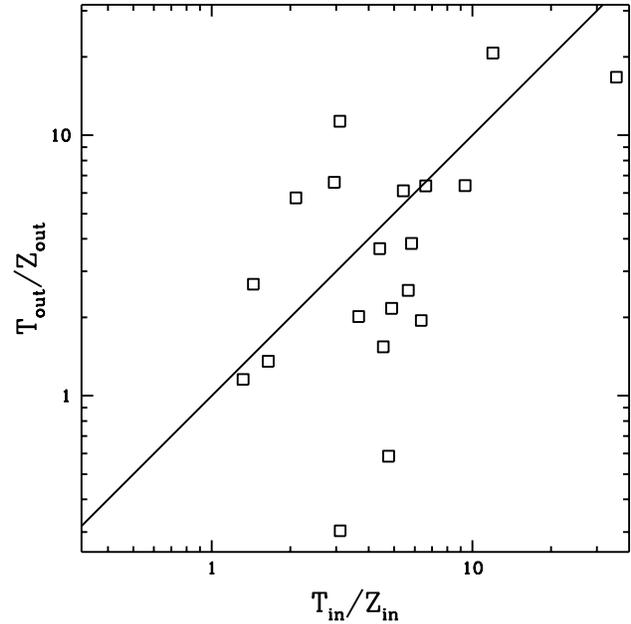,width=0.49\textwidth,angle=0}}
\caption{\label{fig.ratiotest} As Figure \ref{fig.test} except that
the ratios of temperature to abundance are plotted.}
\end{figure}

Finally, in Figure \ref{fig.ratiotest} we plot the ratio of
temperature to abundance for region (out) versus region (in).  If the
Fe-L emission of the MEKAL model is incorrect, then $T/Z$ should tend
to be larger in region (in) if MEKAL predicts too much Fe-L emission
or smaller if MEKAL predicts to little Fe-L emission; i.e. the slope
in the $T/Z$ (in) -- $T/Z$ (out) plane will be negative if the MEKAL
model incorrectly predicts the Fe-L emission. However, no such trend
is found in Figure \ref{fig.ratiotest}. Rather, $T/Z$ for the regions
tend to track each other, albeit with large scatter which is due
mostly to the scatter in the abundances. 

Hence, over 0.5-2 keV the MEKAL code gives very consistent results
when applied to only the Fe-L region (in) and to the region excluding
the Fe-L lines (out). The very sub-solar abundances determined from
the single-temperature fits of the MEKAL model are thus not the result
of errors in the Fe-L portion of the plasma code. As we show in the
next section, these sub-solar abundances arise from fitting a
single-temperature model to an intrinsically multi-temperature
spectrum.

(We have performed the same tests using the Raymond-Smith model. The
results are qualitatively similar to those found with the MEKAL model,
although the scatter in the parameter relationships in regions (in)
and (out) with the Raymond-Smith model is larger. Also $T_{\rm in}$ tends
to exceed $T_{\rm out}$ for the Raymond-Smith model. Although the
Raymond-Smith model, as expected, gives more evidence for problems in
the Fe-L regions than the MEKAL model, the differences do not appear
to be huge. This is supported by the qualitative agreement that we
obtained for the temperatures and abundances of the single-temperature
MEKAL and Raymond-Smith models, though the higher temperatures and
lower abundances of the Raymond-Smith model found in the previous
section are consistent with over-prediction of the Fe-L emission.  As
we discuss below, a large discrepancy between the MEKAL and
Raymond-Smith models is not found for the two-temperature fits as
well.)

\subsection{Two-component fits}
\label{2t}

\renewcommand{\arraystretch}{1.5}

\begin{table*}
\begin{minipage}{175mm}
\caption{Two-Temperature Fits}
\label{tab.2t}
\begin{tabular}{lccccllrrrr}
Name & $N_{\rm H}$ & $T_{\rm C}$ & $T_{\rm H}$ & $Z$ & \multicolumn{2}{c}{$f_{\rm C}/f_{\rm H}$} &
$\Delta\chi^2(N_{\rm H})$ & $\chi^2$ & dof & $\chi^2_{\rm red}$\\ 
     & ($10^{21}$ cm$^{-2}$) & (keV) & (keV) & $(Z_{\sun})$ & (0.5-2
keV) & (0.5-5 keV)\\ 
NGC 499 & $2.0_{-0.4}^{+0.4}$ & $0.65_{-0.02}^{+0.03}$ & $5.7(>3.1)$ &
$0.67_{-0.24}^{+0.53}$ & $11_{-4}^{+7}$ & $4.8_{-1.2}^{+1.9}$ & 32.2 &
231.9 & 200 & 1.16\\ 
NGC 507 & $1.3_{-0.5}^{+0.5}$ & $0.89_{-0.22}^{+0.25}$ &
$1.48_{-0.17}^{+0.29}$ & $0.63_{-0.20}^{+0.32}$ &
$0.32_{-0.23}^{+0.31}$ & $0.25_{-0.18}^{+0.32}$ & 10.4 & 60.8 & 72 &
0.85\\      
NGC 720 & $\cdots$ & $0.57_{-0.05}^{+0.04}$ & $7.4(>3.6)$ &
$0.4(>0.1)$ & $2.9_{-0.9}^{+1.7}$ & $1.4_{-0.3}^{+0.7}$ & $\cdots$ &
42.6 & 45 & 0.95\\
NGC 1332 & $\cdots$ & $0.54_{-0.05}^{+0.04}$ & $14(>7)$ & $0.8(>0.2)$ &
$2.22_{-0.22}^{+0.78}$ & $0.96_{-0.20}^{+0.39}$ & $\cdots$ & 66.9 & 63
& 1.06\\ 
NGC 1399 & $0.9_{-0.3}^{+0.3}$ &  $0.78_{-0.05}^{+0.05}$&
$1.64_{-0.09}^{+0.12}$  & $1.12_{-0.23}^{+0.43}$ &
$0.40_{-0.03}^{+0.05}$ &  $0.31_{-0.03}^{+0.03}$ & 28.8 & 310.9 & 251
& 1.24\\ 
NGC 1404 & $\cdots$ &  $0.60_{-0.02}^{+0.02}$ &  $2.13_{-0.40}^{+0.92}$ &
$0.7_{-0.3}^{+2}$ & $3.9_{-1.3}^{+1.2}$ & $2.7_{-0.8}^{+1.1}$ &
$\cdots$ & 168.8 & 123  & 1.37\\
NGC 1407 & $\cdots$ & $0.78_{-0.04}^{+0.04}$ & $3.9(>2.9)$ &
$0.53_{-0.36}^{+0.98}$ &  $1.59_{-0.33}^{+2.35}$ &
$0.87_{-0.16}^{+0.90}$ & $\cdots$ & 84.5 & 91 & 0.93\\ 
NGC 3923 & $\cdots$ & $0.55_{-0.13}^{+0.08}$ & $4.2(>2.2)$ & $2.0(>0.1)$
& $1.89_{-0.62}^{+0.93}$ & $0.99_{-0.23}^{+0.65}$ & $\cdots$ & 21.8 &
20 & 1.09\\ 
NGC 4374 & $\cdots$ & $0.66_{-0.04}^{+0.05}$ & $5.0(>2.1)$ &
$0.17_{-0.09}^{+0.17}$ & $7.3_{-3.9}^{+19}$ & $3.9_{-1.8}^{+10}$ &
$\cdots$ & 59.3 & 49 & 1.21\\
NGC 4406 & $1.0_{-0.4}^{+0.4}$ & $0.70_{-0.07}^{+0.03}$ &
$1.22_{-0.28}^{+0.30}$ & $0.52_{-0.18}^{+0.22}$ & $2.9_{-1.4}^{+1.7}$
& $2.4_{-1.0}^{+1.2}$ & 11.5 & 179.3 & 183 & 0.98\\
NGC 4472 & $1.2_{-0.3}^{+0.3}$ & $0.76_{-0.03}^{+0.02}$  &
$1.48_{-0.09}^{+0.11}$  & $1.18_{-0.26}^{+0.39}$ &
$1.31_{-0.08}^{+0.11}$ & $1.02_{-0.06}^{+0.07}$ & 56.6 & 260.0 & 235 &
1.11\\ 
NGC 4636 & $1.2_{-0.2}^{+0.3}$ & $0.65_{-0.01}^{+0.02}$ & $>5.3$ &
$0.35_{-0.05}^{+0.08}$ & $42_{-8}^{+2}$ & $14_{-3}^{+2}$ & 46.8 &
206.0 & 135 & 1.53\\
NGC 4649 & $1.5_{-0.5}^{+0.6}$ & $0.75_{-0.04}^{+0.04}$ &
$1.90_{-0.26}^{+0.42}$ & $0.89_{-0.29}^{+0.86}$ &
$2.01_{-0.19}^{+0.34}$ & $1.36_{-0.10}^{+0.17}$ & 21.0 & 131.8 & 96 &
1.37\\  
NGC 5044 & $1.5_{-0.3}^{+0.2}$ & $0.70_{-0.03}^{+0.02}$ &
$1.20_{-0.05}^{+0.05}$ & $0.62_{-0.08}^{+0.11}$ &
$0.85_{-0.08}^{+0.09}$ & $0.72_{-0.06}^{+0.08}$ & 82.0 & 193.9 & 158 &
1.23\\ 
NGC 5846 & $1.3_{-0.4}^{+0.5}$ & $0.63_{-0.04}^{+0.03}$ &
$1.32_{-0.21}^{+0.18}$ & $0.55_{-0.19}^{+0.29}$ & $2.8_{-0.7}^{+0.9}$
& $2.2_{-0.4}^{+0.6}$ & 13.6 & 92.8 & 104 & 0.89\\
NGC 6876 & $\cdots$ & $0.90_{-0.14}^{+0.14}$ & $3.0(>1.5)$ &
$0.27_{-0.17}^{+0.66}$ & $3.5_{-2.4}^{+29}$ & $2.0_{-1.1}^{+13}$ &
$\cdots$ &  21.0 & 16 & 1.31\\
NGC 7619 & $3.4_{-1.0}^{+1.0}$ & $0.65_{-0.05}^{+0.05}$ & $3.1(>1.8)$ &
$1.2(>0.3)$ & $9.3_{-4.1}^{+16.1}$ & $4.3_{-1.2}^{+3.5}$ & 31.3 & 82.6
& 95 & 0.87\\ 
NGC 7626 & $\cdots$ & $0.69_{-0.06}^{+0.07}$ & $3.3_{-1.0}^{+1.6}$ &
$2.0(>0.4)$ & $1.6_{-0.4}^{+0.6}$ & $1.1_{-0.4}^{+0.2}$ & $\cdots$ & 
52.8 & 49 & 1.08\\  
IC 1459 & $\cdots$ & $0.66_{-0.10}^{+0.09}$ & $7.8_{-3.0}^{+16}$ &
$0.14_{-0.09}^{+0.67}$ &  $0.36_{-0.01}^{+0.67}$ &
$0.22_{-0.03}^{+0.26}$ & $\cdots$ & 58.9 & 52 & 1.13\\
IC 4296 & $\cdots$ & $0.75_{-0.04}^{+0.05}$ & $5.0_{-1.2}^{+1.3}$ &
$2.8(>0.9)$ & $0.94_{-0.13}^{+0.47}$ & $0.49_{-0.05}^{+0.32}$ &
$\cdots$ & 86.7 & 88 & 0.99\\

\end{tabular}
\medskip

Results of fitting 2 MEKAL models each modified by the same
photo-electric absorption.  If allowing $N_{\rm H}$ to be free significantly
improved the fit, we give the best-fit value of $N_{\rm H}$ and the change
in $\chi^2$ denoted by $\Delta\chi^2(N_{\rm H})$.  The ``$\cdots$'' indicate
that letting $N_{\rm H}$ be free did not significantly improve the fits. The
abundances, $Z$, are tied together in the fits. The best-fit values
and their 90\% confidence limits on one interesting parameter are
given. $f_{\rm C}/f_{\rm H}$ is the flux of the absorbed cold component divided by
that of the absorbed hot component.

\end{minipage}
\end{table*}

\renewcommand{\arraystretch}{1.0}

\renewcommand{\arraystretch}{1.5}

\begin{table*}
\begin{minipage}{170mm}
\caption{Three-Temperature Fits}
\label{tab.3t}
\begin{tabular}{lcccccrrrrr}
Name & $N_{\rm H}$ & $T_1$ & $T_2$ & $T_3$ & $Z$ &
\multicolumn{2}{c}{$f_1:f_2:f_3$} & $\chi^2$ & dof & $\chi^2_{\rm red}$\\
& ($10^{-21}$ cm$^{-2}$) & (keV) & (keV) & (keV) & $(Z_{\sun})$ &
(0.5-2 keV) & (0.5-5 keV)\\
NGC 4472 & $1.3_{-0.3}^{+0.3}$ & $0.73_{-0.03}^{+0.03}$ & $1.34_{-0.08}^{+0.10}$ 
& $>6$ & $1.19_{-0.28}^{+0.46}$ & $21:17:1$ & $7.0:7.1:1$
& 246.4 & 230 & 1.07\\ 
NGC 4636 & $1.4_{-0.4}^{+0.5}$ & $0.53_{-0.17}^{+0.10}$ & $0.73_{-0.07}^{+0.20}$
& $8(>3)$ &  $0.47_{-0.12}^{+0.17}$ & $15:12:1$ & $6.1:5.0:1$ &
181.7 & 132 & 1.38\\
NGC 4649 & $2.0_{-0.7}^{+0.9}$ & $0.67_{-0.08}^{+0.07}$ &
$1.15_{-0.22}^{+0.21}$ & $>5$ & $0.88_{-0.33}^{+0.41}$ & $9.7:7.3:1$
& $3.1:2.7:1$ & 110.0 & 93 & 1.18\\
\end{tabular}
\medskip

Results of fitting 3 MEKAL models each modified by the same
photo-electric absorption for three galaxies where the fit is at least
marginally improved. The abundances, $Z$, are tied together in the
fits. The best-fit values and their 90\% confidence limits on one
interesting parameter are given for the column densities, temperatures
and abundances.  The best-fitting ratio of the fluxes of component 1
to component 3 and the ratio of component 2 to component 3 are given
by $f_1:f_2:f_3$.

\end{minipage}
\end{table*}

\renewcommand{\arraystretch}{1.0}

\begin{figure*}
\parbox{0.49\textwidth}{
\centerline{\psfig{figure=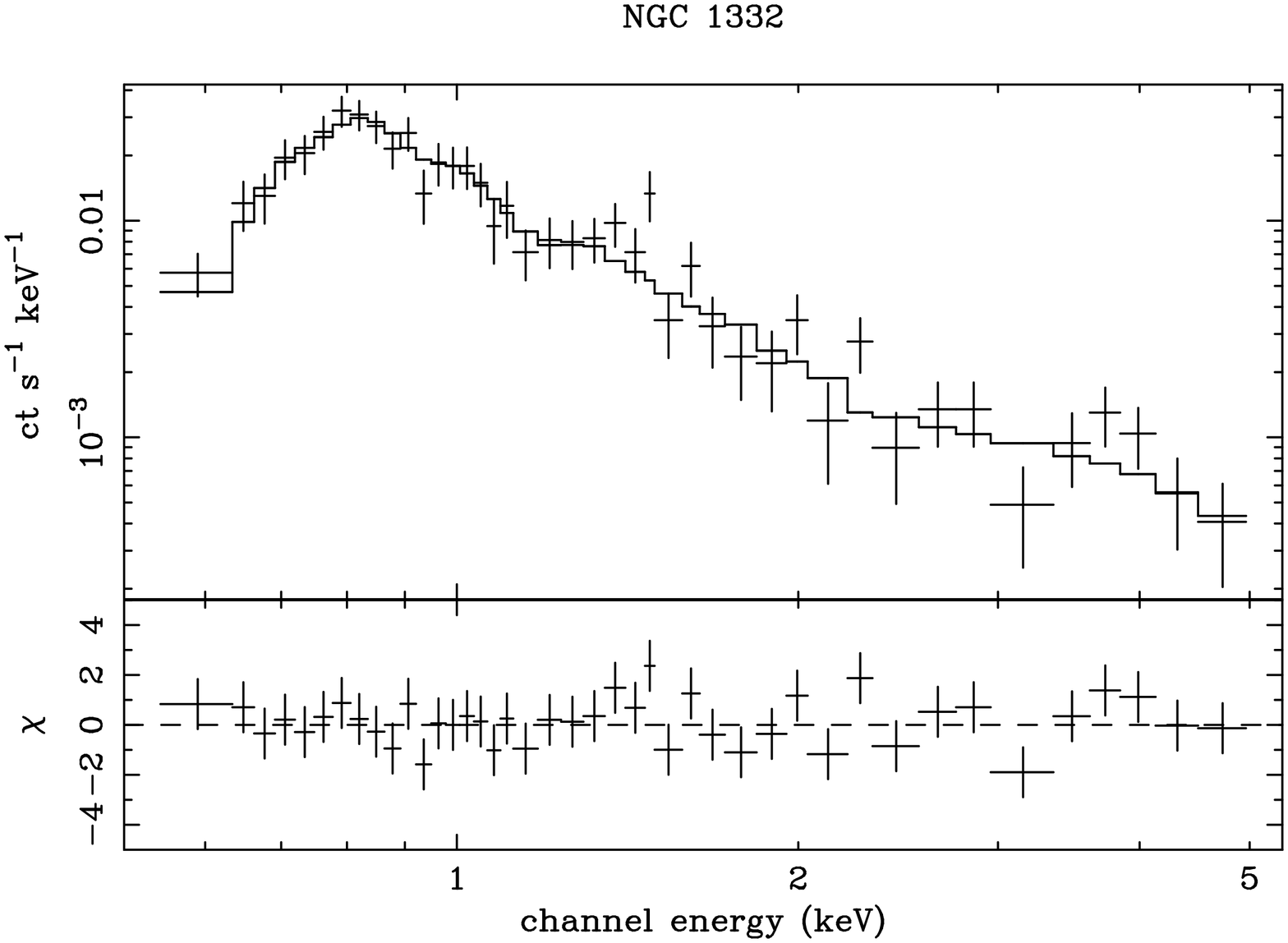,angle=-90,height=0.3\textheight}}
}
\parbox{0.49\textwidth}{
\centerline{\psfig{figure=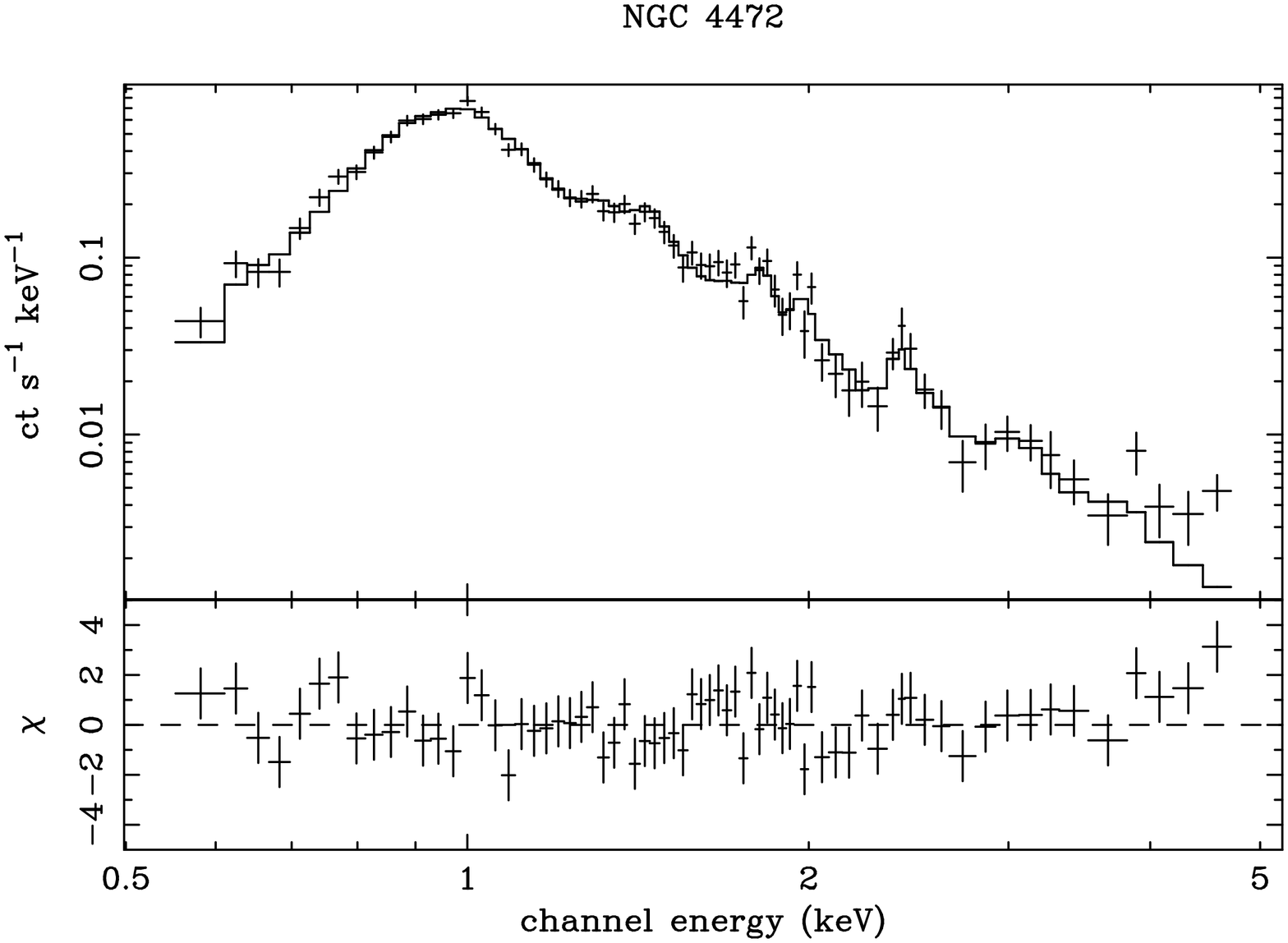,angle=-90,height=0.3\textheight}}
}
\vspace{-0.2cm}
\parbox{0.49\textwidth}{
\centerline{\psfig{figure=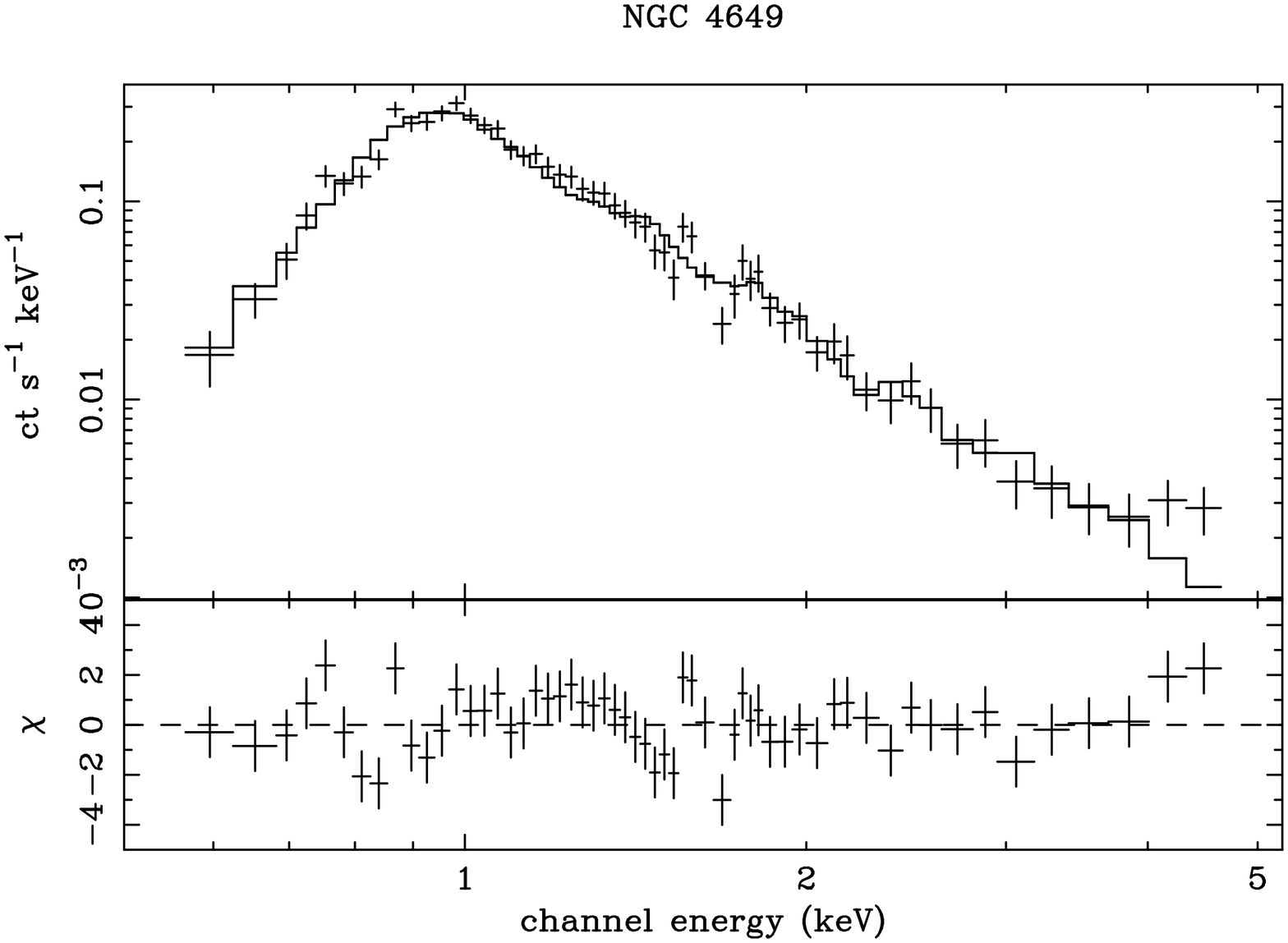,angle=-90,height=0.3\textheight}}
}
\parbox{0.49\textwidth}{
\centerline{\psfig{figure=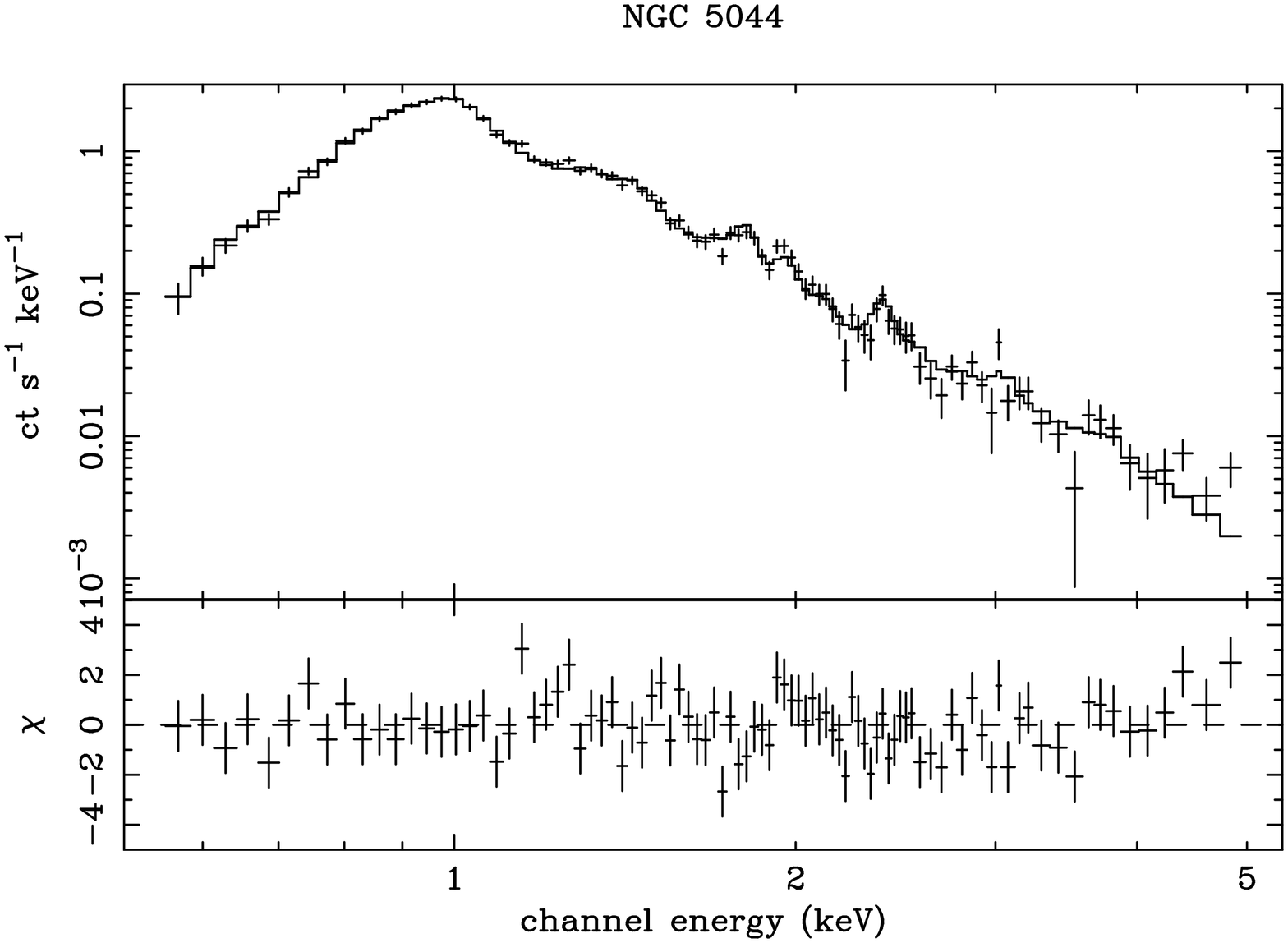,angle=-90,height=0.3\textheight}}
}
\caption{\label{fig.2t} As Figure \ref{fig.1t} but the spectra are fit
with absorbed two-temperature (MEKAL) models.}
\end{figure*}

Since most of the single-temperature fits with the MEKAL model clearly
provide unacceptable fits to the {\sl ASCA} spectra, we examined
whether the fits would improve significantly upon adding another
temperature component. Our procedure is to begin with the best-fitted
single-temperature model with Galactic $N_{\rm H}$ and then add
another absorbed temperature component with $N_{\rm H}$ and $Z$ tied
to the first component. The normalization of the second component is
initially set to zero. After fitting this two-temperature model we
then allowed $N_{\rm H}$ to be free to see if the fit could be further
improved.  

The process of fitting these two component models to the {\sl ASCA}
spectra required some care to insure that the global $\chi^2$ minimum
was achieved by XSPEC for the galaxies having lowest S/N in our
sample; i.e. for approximately half of the galaxies.  Typically, upon
initially fitting the two-temperature model to these lower S/N
galaxies one finds that the first component, let us call it the cold
component with temperature $T_{\rm C}$, has best-fit $T_{\rm C}\la 1$
keV similar to what was obtained with the single-temperature fits. The
temperature of the second component, let us call it the hot component
with temperature $T_{\rm H}$, typically has best-fit $T_{\rm H}\ga 10$
keV. Also, the abundances tended to differ insignificantly from their
single-temperature values.

We found, however, that for many of these galaxies these fits were not
the global minima. By taking these best-fit results and then initially
setting $T_{\rm H}$ to some relatively small value (e.g. $1-2$ keV)
and the abundances to some relatively large value $Z\sim
(1-2)Z_{\sun}$ and then fitting again we found that a deeper minimum
was usually obtained. The results of this new deeper minimum were
generally smaller values of $T_{\rm H}$ and larger values of $Z$,
though in most cases the 90 per cent confidence regions overlapped
with those of the previous minimum.

For example, the initial fit for IC 4296 yielded ($T_{\rm C}=0.81$
keV, $T_{\rm H}=18.4$ keV, $Z=0.20Z_{\sun}$, $\chi^2=93.8$). By then
resetting $T_{\rm H}$ and $Z$ to small and large values respectively,
we obtained (after performing this procedure twice) best fit values
($T_{\rm C} = 0.75$ keV, $T_{\rm H} = 5.0$ keV, $Z=2.8Z_{\sun}$,
$\chi^2=86.7$) as given in Table \ref{tab.2t}: the change in $\chi^2$
is greater than 2.71. The other galaxies showed a smaller
effect. Initial fits to NGC 720 gave ($T_{\rm C}=0.57$ keV, $T_{\rm
H}=41.4$ keV, $Z=0.18Z_{\sun}$, $\chi^2=43.7$). The results listed in
Table \ref{tab.2t} show that the change in $\chi^2$ is only 1.1, but
the best-fit parameters and their confidence regions (though
overlapping) are qualitatively different; i.e. the abundance in the
deeper minimum is unbounded from above unlike that in the shallower
minimum.

In Table \ref{tab.2t} we list the results of fitting two absorbed
MEKAL models to each galaxy in the sample.  In Figure \ref{fig.2t} we
plot the two-temperature fits corresponding to the same galaxies in
Figure \ref{fig.1t}. In all but a few cases the improvement in the
fits is substantial and the values of reduced $\chi^2$ are brought to
approximately 1.0.  (We found that untying the abundances and/or
absorption of the components did not significantly improve the fits.)

As expected, the galaxies like NGC 507, NGC 4374, and NGC 4406, which
already had reduced $\chi^2\approx 1.0$ from fits to a single MEKAL
model, generally showed the smallest improvement when another
component was added.  For NGC 1404 the poor fit is mostly due to
residuals near 1.8 keV in the observation sequence 80039000. The
origin of these residuals is unclear (they are not seen in the
spectrum of NGC 1399 from the same sequence), but if we exclude this
sequence we obtain a reduced $\chi^2\sim 1.0$ for the
single-temperature model and thus, like NGC 507, NGC 4374, and NGC
4406, a single temperature appears to produce the emission without
great (formal) need for another component.

For NGC 6876, which has the poorest S/N of the sample, letting $N_{\rm
H}$ be free substantially improved the fit, $\chi^2\approx 16.5$, but
the resulting fitted parameters were very poorly constrained. (Because
of the low S/N we were unable to obtain a definite best-fit result
because the $\chi^2$ space contained many small isolated minima of
similar depth.) The 90 per cent confidence regions, however, include
the parameter space obtained when $N_{\rm H}$ is held at the galactic
value, and thus we have presented those results in Table \ref{tab.2t}.

As expected (see \S \ref{obs}), when fitting the SIS+GIS data we
found, as with the single-temperature fits of the previous section,
that the results differed negligibly from the SIS results alone. We
shall henceforth focus on the SIS results.

\subsubsection{Temperatures}

The effect of adding the second temperature component can be divided
roughly into two classes.  For the galaxies with the best S/N and
typically the largest $L_{\rm x}/L_{\rm B}$ the two temperature
components essentially bracket the single-temperature value with
$T_{\rm C}\sim 0.5-1$ keV and $T_{\rm H}\sim 1-2$ keV. These
temperatures are consistent with emission from hot gas.  For the
galaxies having large S/N and large $L_{\rm x}/L_{\rm B}$ but $T_{\rm
H}>2$ keV (NGC 499, NGC 4636, NGC 7619), the cold component dominates
the X-ray emission. Hence, the high S/N, high $L_{\rm x}/L_{\rm B}$
galaxies either have no evidence for emission from a hard component
with $T_{\rm H}\ga 5$ keV or allow for only a small fraction $(\la
10\%)$ of the total flux to reside in such a component. Rather,
multiple phases in the hot gas are indicated by the two-temperature
(and three-temperature, see below) fits.  The fits to most of these
galaxies are also significantly improved by allowing for excess
absorption. In \S \ref{nh} we discuss in detail the evidence and
implications for this excess absorbing material. (We reiterate that
the fits are not significantly improved by allowing for individually
fitted abundances for the cold and hot components, or by letting the
$\alpha$-process elements be fitted separately from Fe.)

The other galaxies, those with lowest S/N and smallest $L_{\rm
x}/L_{\rm B}$, typically have $T_{\rm C}\sim 0.5-1$ keV (similar to
the single-temperature value) and $T_{\rm H}\sim 5-10$ keV. The cold
component is consistent with emission from hot gas while the hotter
component has temperatures consistent with that expected from discrete
sources (e.g. Kim et al. 1992); i.e. this putative discrete component
is found to be most important in galaxies that tend to have the lowest
$L_{\rm x}/L_{\rm B}$ ratios in our sample as expected \cite{cft}.
Typically the emission from the cold component exceeds that of the hot
component in the soft energy range, 0.5-2 keV, while the emission from
each component is roughly equal over the full 0.5-5 keV band. IC 1459
is the exception as its hot component dominates which again suggests
that its emission is qualitatively different from others in our sample
and may indicate the emission from the AGN.

For the galaxies with $T_{\rm H}\sim 3-5$ keV having large $L_{\rm x}/L_{\rm B}$ but
average S/N in our sample, it is possible that higher S/N data will
split $T_{\rm H}$ into another soft component and a hard component. (This
may even happen for the lowest $L_{\rm x}/L_{\rm B}$ galaxies.)  Interpretation of
the nature of the hotter component in these galaxies must await better
quality data.

\begin{figure*}
\parbox{0.49\textwidth}{
\centerline{\psfig{figure=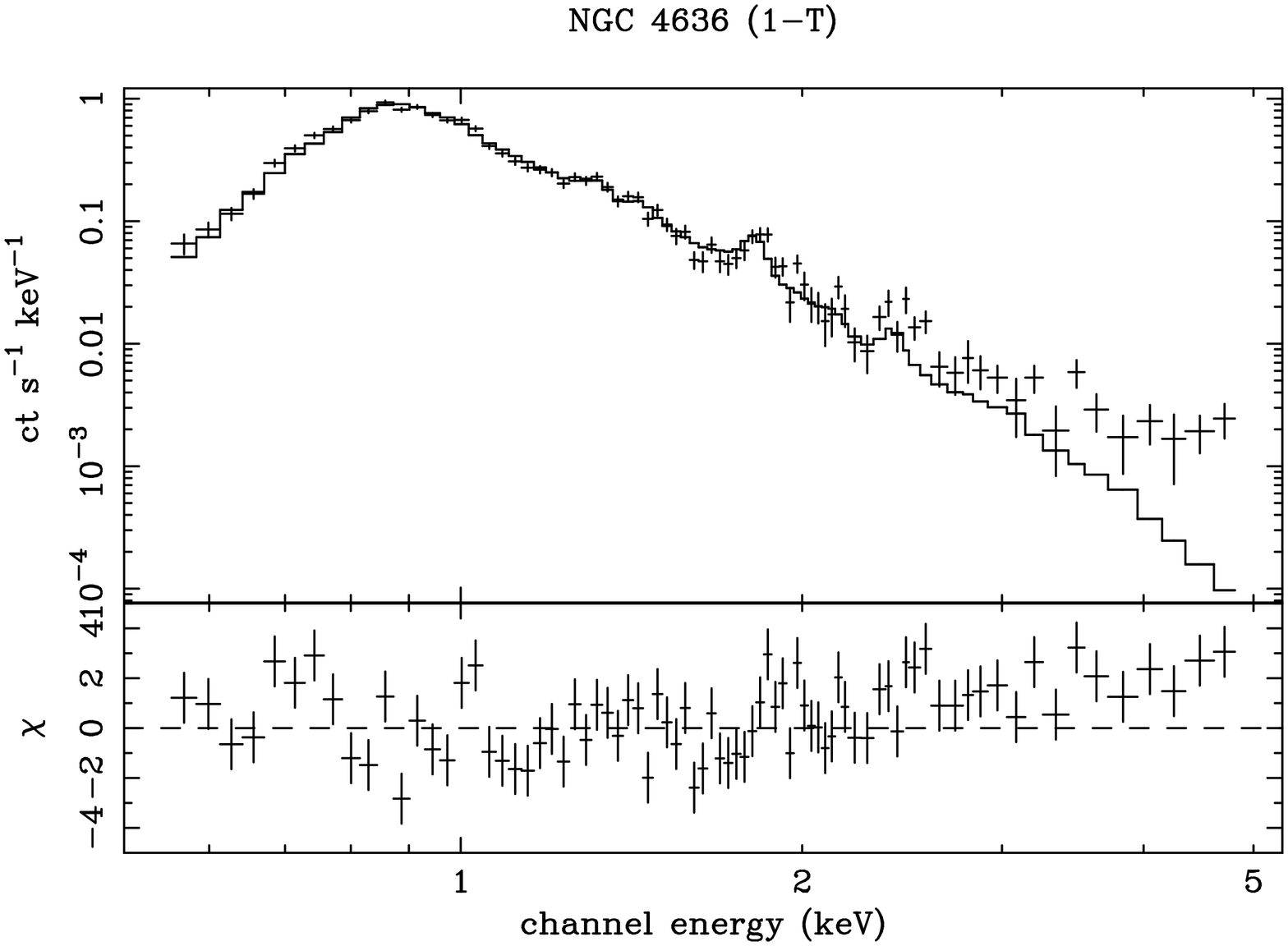,angle=-90,height=0.3\textheight}}
}
\parbox{0.49\textwidth}{
\centerline{\psfig{figure=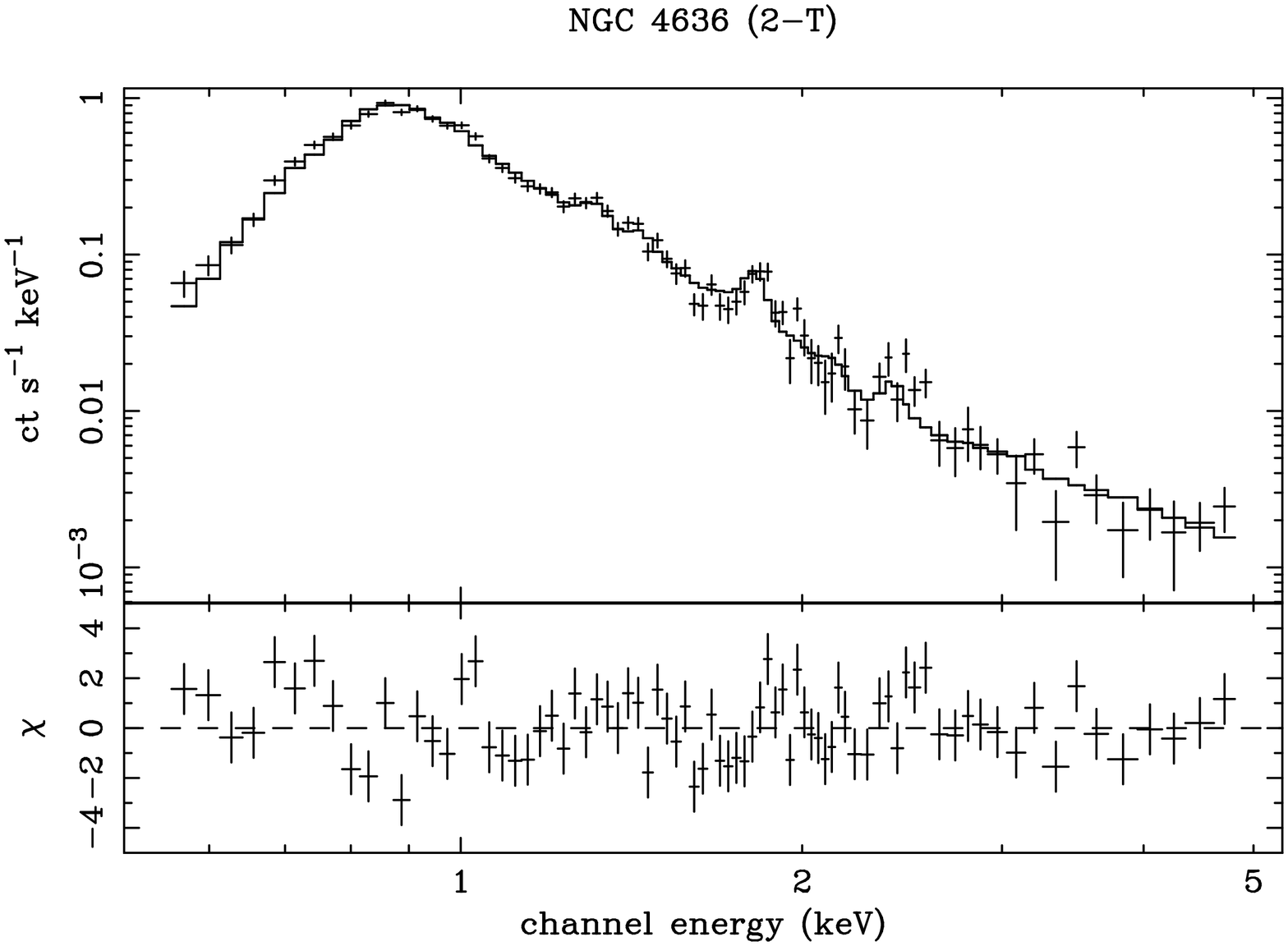,angle=-90,height=0.3\textheight}}
}
\vspace{-0.2cm}
\parbox{0.49\textwidth}{
\centerline{\psfig{figure=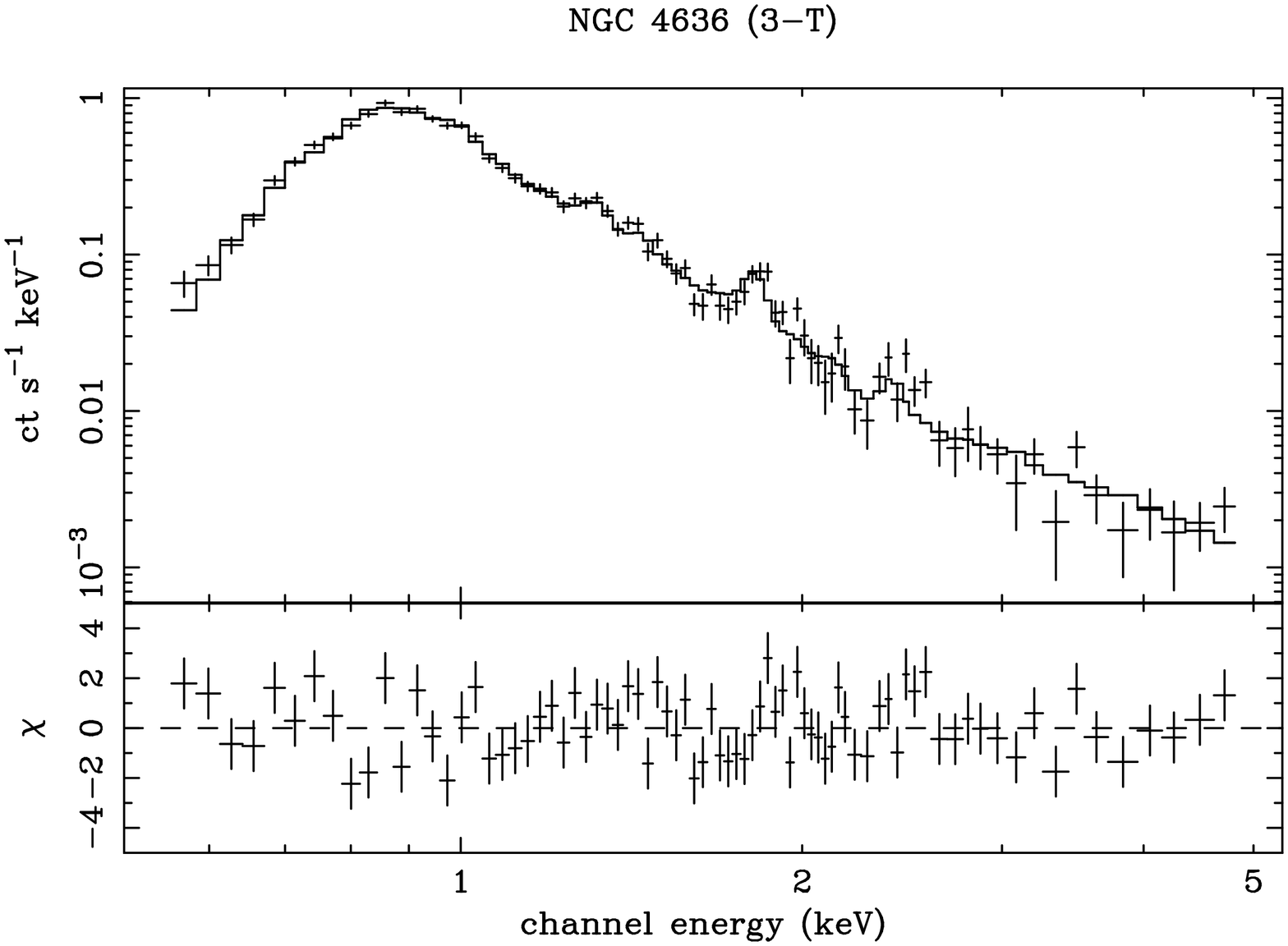,angle=-90,height=0.3\textheight}}
}
\caption{\label{fig.n4636} Single-temperature (1-T), two-temperature
(2-T), and three-temperature (3-T) (MEKAL) models fit to NGC
4636. Only the SIS0 data is shown.}
\end{figure*}

NGC 4636 has the worst formal fit of the galaxies in our sample which
does not have any obvious anomalous features in its spectrum (see
Figure \ref{fig.n4636}). Untying the absorptions and abundances of the
components did not improve the fits. We tried grouping the
$\alpha$-process elements (O, Ne, Mg, Si, and S) separately from Fe
and the rest to no avail. The best improvement in $\chi^2$ was
obtained by adding a third temperature component (see Table
\ref{tab.3t} and Figure \ref{fig.n4636}).

The fit was not improved greatly and so there may be something special
to NGC 4636 inhibiting a good fit. A new 200ks observation of NGC 4636
shows the same anomalous spectrum indicating that the features are
real and not the result of a spurious observation (Buote, Fabian, \&
Canizares 1997, in preparation).  The addition of the third
temperature component split $T_{\rm C}$ into two roughly equal
components with temperatures between 0.5-1 keV. The flux of the
high-temperature component remains dwarfed by the soft components.
The substantial residuals for $E<1$ keV and the fact that NGC 4636 has
a low value of $T_{\rm C}$ for its X-ray luminosity (see Figure
\ref{fig.lxtx} in \S \ref{lxlb}) suggests the presence of additional
cold phases.

Essentially all of the other galaxies are not affected significantly
by adding another temperature component, although NGC 4472 and
especially NGC 4649 show some improvement (see Table \ref{tab.3t}). In
these two cases the added component has a temperature with no upper
bound and may indicate the emission from discrete sources in those
galaxies. Again, the softer components dominate the hard component.

Fitting two Raymond-Smith models to the galaxy spectra gives $\chi^2$
values that are qualitatively similar to those obtained from the MEKAL
models in many cases, but there are notable differences.  Generally,
the reduced $\chi^2$ values of the Raymond-Smith fits are larger,
$(\chi^2_{\rm red})^{\rm RS}\sim(\chi^2_{\rm red})^{\rm
MEKAL}+0.2$. However, the fits to NGC 1399 $(\chi^2_{\rm red}=1.43)$
and NGC 4472 $(\chi^2_{\rm red}=1.47)$ are noticeably poorer than the
MEKAL fits while NGC 4636 $(\chi^2_{\rm red}=2.65)$ and NGC 4472
$(\chi^2_{\rm red}=1.96)$ have dramatically worse fits. These galaxies
coincidentally have the best S/N (and among the largest $L_{\rm
x}/L_{\rm B}$) in the sample.

When $N_{\rm H}$ was allowed to be free the fits with the
Raymond-Smith model were not improved as much as were fits with the
MEKAL model.  Only NGC 507, NGC 4649, NGC 5044, NGC 6876, and NGC 7619
showed significant improvement but with a lesser change in $\chi^2$
than found with the MEKAL model.

The fits to the galaxies using the Raymond-Smith code also gave
qualitatively comparable temperatures to those determined with MEKAL,
although both $T_{\rm C}$ and $T_{\rm H}$ tend to be larger for
Raymond-Smith\footnote{This is consistent with the Raymond-Smith model
predicting too much Fe-L emission for a particular $T$ and $Z$ -- see
\S \ref{abun.1t}.}. Typically, $T_{\rm C}^{\rm RS} \sim T_{\rm C}^{\rm
MEKAL} + 0.2$. For galaxies where $T_{\rm H}\sim 5$ keV as determined
from the MEKAL model, the Raymond-Smith fits tend to give $T_{\rm H}>
10$ keV; for many of these only lower limits could be found.  We found
that the Raymond-Smith fits gave $T_{\rm H}<2$ keV for only NGC 1404,
NGC 4472, NGC 5044, and NGC 5846. Interestingly, when $N_{\rm H}$ is
allowed to be free for NGC 4472 the best-fit value of $T_{\rm H}$
increases without bound, although the quality of the fit is not
significantly improved. (The abundances in these two cases are quite
different as we mention below.)  Hence, the nature of the emission of
the second temperature component, be it due to another phase of the
hot gas or to discrete sources, is dependent on the plasma code;
i.e. MEKAL tends to give smaller $T_{\rm H}$ and thus more indication
of additional phases of the hot gas than Raymond-Smith.

Ten of the galaxies in our sample have been previously analyzed by
Matsumoto et al. \shortcite{mat} using {\sl ASCA} data (see \S
\ref{intro}). These authors fit a two-component model to the SIS and
GIS spectra where each component is modified by the same variable
absorption, and the abundances of each component are tied together.
For IC 1459, IC 4296, NGC 499, NGC 720, NGC 4374, and NGC 4636, where
we found $T_{\rm C}\sim 0.5-1$ keV and $T_{\rm H}\ga 5$ keV, Matsumoto
et al. obtained $T_{\rm C}$ and reduced $\chi^2$ values (and $N_{\rm
H}$) in good agreement with our results -- whether we used two MEKAL
models or Raymond-Smith models. (We have different reduced $\chi^2$
values for NGC 507 because we excluded its anomalous SIS1 data -- see
beginning of \S \ref{fit}.)  For NGC 4406, NGC 4472, and NGC 4649
Matsumoto et al. obtained comparable quality fits to us without
requiring a second hot phase. As mentioned above, this is due to
fitting Raymond-Smith models instead of MEKAL models.

We also obtained essentially the same $T_{\rm C}$ and $T_{\rm H}$ of
Buote \& Canizares \shortcite{BC97} who fitted two Raymond-Smith
components to NGC 720 and NGC 1332.  The unbounded $T_{\rm H}$ (and
low abundance) obtained by Buote \& Canizares using two Raymond-Smith
models agrees with our findings using the same models.  Our
Raymond-Smith fits for NGC 4374, NGC 4406, NGC 4472, and NGC 4636 give
$T_{\rm C}$ and $T_{\rm H}$ in good agreement with the results of
Matsushita et al. \shortcite{matsu} who fitted two Raymond-Smith
components to the GIS data.

Our results for NGC 5044 appear to be consistent with those from
Fukazawa et al. \shortcite{fuk} who fitted single-temperature models
to {\sl ASCA} SIS and GIS data in the regions $R<2\arcmin$ and
$4\arcmin<R<6\arcmin$ and found $\chi^2_{\rm red}>1.42$ for a suite of
4 plasma codes.  Since the residuals of the fit of their
single-temperature Raymond-Smith model (see their Figure 4) are very
consistent with our findings (e.g. with the MEKAL model in Figure
\ref{fig.1t}), a two-temperature model with both temperatures less
than 2 keV would appear to be required for an acceptable fit to their
spectra.

\subsubsection{Abundances}

The metal abundances obtained from the two-temperature (and
three-temperature) MEKAL models are substantially larger than those
obtained from the single-temperature models.  For 18 out of 20
galaxies we find that the best-fit metal abundances for the
two-temperature models have $Z > 0.4Z_{\sun}$.  The mean and standard
deviation of the abundances for the entire sample are $\langle
Z\rangle = 0.9\pm 0.7$ $Z_{\sun}$. The two galaxies (IC 1459 and NGC
4374) having very subsolar abundances also have among the lowest S/N
spectra in our sample\footnote{We note that it would be better to
quote a mean abundance weighted by the uncertainties in the individual
measurements. However, the error bars are highly asymmetric and the
sizes of the errors are highly correlated with the measured value of
$Z$. Thus, a simple weighting by standard errors is
inappropriate.}.

This systematic increase of $Z$ derived for the two-temperature models
with respect to single-temperature models occurs because of an
intrinsic bias specific to X-ray spectral fitting.  As noted in \S 2.2
of Buote \& Canizares \shortcite{BC94} regarding the analysis of the
{\sl ROSAT} PSPC spectrum of NGC 720 with the Raymond-Smith code, when
a galaxy spectrum which is intrinsically described by a
two-temperature solar abundance model is fitted with a
single-temperature model, the fitted abundance will necessarily have a
value that is very subsolar; i.e. the very subsolar abundances are an
artificial systematic bias of fitting single-temperature models to
intrinsically multi-temperature spectra. We verified this result using
simulated two-temperature {\em ASCA} spectra with MEKAL models for the
highest S/N galaxies in our present sample\footnote{Some {\sl ROSAT}
studies cautioned that poor quality spectra simply could not
distinguish between single-temperature models with very sub-solar
abundances and two-temperature models with solar abundances (e.g.
Trinchieri et al. 1994) owing to the larger number of free parameters
in the latter case.}. (We mention that for the galaxies where excess
absorption improved the fits, when the absorption is fixed to its
Galactic value the abundances are usually about the same as for the
fits allowing for excess absorption, though in some cases the
abundances are considerably larger; e.g. for NGC 4649 we obtain
$Z=0.89Z_{\sun}$ when allowing for excess absorption but
$Z=1.24Z_{\sun}$ for Galactic $N_{\rm H}$. Thus the relatively large
abundances are not the result of the (in some cases) large required
$N_{\rm H}$ suppressing the excess line emission at lower energies.)

Hence, the very subsolar abundances obtained by previous studies with
{\sl ROSAT} (e.g. Davis \& White 1996) and {\sl ASCA} (e.g. Arimoto
et al. 1997) which fitted single-temperature models to the X-ray
spectra of ellipticals appear to be a fitting artifact. However, even
recent studies with {\sl ASCA} which fitted two-temperature models
obtained typical abundances that are significantly less than
solar. For example, Matsushita et al. \shortcite{matsu} analyzed a
small sample of early-type galaxies and obtained $Z\approx
0.4Z_{\sun}$ for NGC 4406, NGC 4472, NGC 4636, and $Z\approx
0.1Z_{\sun}$ for NGC 4374 by fitting two Raymond-Smith components to
GIS data. Our Raymond-Smith results produce similar values except for
NGC 4472 where we obtain $Z=1.3Z_{\sun}$.  We obtain qualitatively
similar results for these galaxies with fits to two MEKAL models. If
we let $N_{\rm H}$ be free for the fit to NGC 4472 with two
Raymond-Smith models, then we obtain $Z=0.5Z_{\sun}$ in much better
agreement with Matsushita et al. (although the fit is not
improved significantly).

These four galaxies, however, have somewhat smaller abundances than
our entire sample. For the fits with two Raymond-Smith models we
obtain for the whole sample an abundance with mean and standard
deviation, $\langle Z\rangle = 0.7\pm 0.6$ $Z_{\sun}$. This value is
systematically smaller than obtained from the fits with two MEKAL
models, but the relatively small systematic difference is consistent
with our findings in \S \ref{abun.1t}.

However, these abundances are still systematically larger than the
abundances $(Z\la 0.4Z_{\sun})$ obtained by Matsumoto et
al. \shortcite{mat} who analyzed a sample of 12 early-type galaxies
(see \S \ref{intro}). The disagreement with our results probably is
due to the different modeling procedures. That is, although Matsumoto
et al. use a variable temperature Raymond-Smith model for their soft
component, they use a fixed, 12 keV, Bremsstrahlung temperature for
the hard component. (And they let $N_{\rm H}$ be free in all cases,
regardless of whether the fit demands that it be different from the
Galactic values.) We explained near the beginning of \S \ref{2t} that
larger values of $T_{\rm H}$ tend to imply smaller values of $Z$,
especially for the galaxies having spectra with the lowest
S/N. Considering that Matsumoto et al. generally use larger extraction
apertures than us (particularly for the SIS1) their spectra have lower
S/N and thus these effects become more pronounced.

Thus, the modeling differences and lower S/N data likely account for
the differences between our present study and those of Matsumoto et
al. (The updated {\sl ASCA} calibration used in our paper also
contributes to these differences.) Since the MEKAL model is more
accurate than Raymond-Smith, and we believe there is no reason to bias
the fits with a preconceived hard component required to be present in
the same form in all galaxies, and our data have generally higher S/N,
the approximately solar abundances found in this paper would appear to
be favored.  Possible abundance gradients do not resolve our
discrepancy with previous studies because we obtain approximately
solar abundances even for some of the large S/N galaxies in our sample
which have large extraction radii (e.g. NGC 1399, NGC 4472).
Approximately solar abundances are consistent with multi-phase models
for the evolution of the hot gas in ellipticals (e.g. Fujita,
Fukumoto, \& Okoshi 1997) but are still inconsistent with the highly
super-solar abundances predicted by standard enrichment models (e.g,
Ciotti et al. 1991; also see Arimoto et al. 1997).

\subsubsection{Cooling flows}

We also fitted the galaxy spectra to a two-component model consisting
of a cooling-flow model for one component and a MEKAL model for the
other. The absorption and abundances were handled as with the two
MEKAL models above. In all cases we found that these models gave fits
of essentially the same quality as obtained with the 2 MEKAL
models. Generally, the fitted temperatures and abundances behaved
similarly as well. For galaxies with both temperature components less
than about 2 keV the MEKAL component can be thought to represent
either the ambient gas phase or the gravitational work done on the
flow, and the cooling-flow component the distributed mass drop-out
term. It should be emphasized that $T_{\rm CF}$ is the maximum
temperature from which the gas cools (i.e. not the emission-weighted
temperature).

For example, we found for NGC 5044 the temperature of the cooling-flow
component to be, $T_{\rm CF}=1.6$ keV, that of the MEKAL component to
be, $T_M=0.94$ keV, and the abundance to be $Z=0.94Z_{\sun}$; as with
the single-temperature fits, the abundances obtained with cooling-flow
models are larger.  The flux of the cooling-flow component is
dominant, $f_{\rm CF}/f_M=3.6$ (0.5-2 keV). The inferred mass
deposition rate, $\dot{M}_{\rm gas}=47M_{\sun}$ yr$^{-1}$, is less
than that obtained in the single-component case since the flux is now
shared with the other component.

\subsubsection{Luminosities}
\label{lxlb}

\begin{table*}
\begin{minipage}{128mm}
\caption{Luminosities}
\label{tab.lum}
\begin{tabular}{lcccccc}
Name &  \multicolumn{2}{c}{$\log_{10} L_{\rm x}^C$} &
\multicolumn{2}{c}{$\log_{10} L_{\rm x}^H$} & \multicolumn{2}{c}{$\log_{10}
L_{\rm x}/L_{\rm B}$}\\ 
& (0.5-2 keV) & (0.5-5 keV) & (0.5-2 keV) & (0.5-5 keV) & (0.5-2 keV)
& (0.5-5 keV)  \\  
NGC 499  &   42.55 & 42.57&  41.39 & 41.70 & -1.16 & -1.12\\
NGC 507  &   42.19 & 42.21&  42.66 & 42.76 & -1.40 & -1.32\\
NGC 720  &   40.76 & 40.77&  40.30 & 40.58 & -2.58 & -2.48\\
NGC 1332 &   40.43 & 40.44&  40.07 & 40.43 & -2.70 & -2.55\\
NGC 1399 &   41.30 & 41.32&  41.66 & 41.77 & -1.65 & -1.56\\
NGC 1404 &   41.23 & 41.24&  40.64 & 40.80 & -2.02 & -1.97\\
NGC 1407 &   40.97 & 40.99&  40.75 & 41.01 & -2.43 & -2.31\\
NGC 3923 &   40.89 & 40.90&  40.58 & 40.83 & -2.76 & -2.65\\
NGC 4374 &   41.04 & 41.06&  40.17 & 40.45 & -2.57 & -2.51\\
NGC 4406 &   41.78 & 41.79&  41.29 & 41.37 & -1.86 & -1.83\\
NGC 4472 &   41.59 & 41.61&  41.43 & 41.54 & -2.22 & -2.17\\
NGC 4636 &   41.97 & 41.98&  40.28 & 40.69 & -1.60 & -1.58\\
NGC 4649 &   41.41 & 41.42&  41.06 & 41.20 & -2.26 & -2.20\\
NGC 5044 &   42.80 & 42.81&  42.82 & 42.89 & -0.64 & -0.60\\
NGC 5846 &   41.96 & 41.97&  41.46 & 41.54 & -1.69 & -1.66\\
NGC 6876 &   41.51 & 41.55&  41.00 & 41.23 & -2.03 & -1.94\\
NGC 7619 &   42.25 & 42.26&  41.09 & 41.31 & -1.51 & -1.49\\
NGC 7626 &   41.06 & 41.07&  40.82 & 41.05 & -2.44 & -2.34\\
IC 1459 &   40.47 & 40.49&  40.71 & 41.03 & -2.62 & -2.39\\
IC 4296 &   41.53 & 42.00&  41.54 & 41.81 & -2.29 & -1.90\\
\end{tabular}

\medskip

Intrinsic (unabsorbed) luminosities of the cold component $(L_{\rm x}^C)$
and the hot component $(L_{\rm x}^H)$ in erg s$^{-1}$ corresponding to the
two-temperature models of Table \ref{tab.2t}.  We used the distances
from Faber et al. \shortcite{7s} as reported in Eskridge, Fabbiano, \&
Kim \shortcite{efk} for $H_0=50$ km/s/Mpc.

\end{minipage}
\end{table*}

\begin{figure*}
\parbox{0.49\textwidth}{
\centerline{\psfig{figure=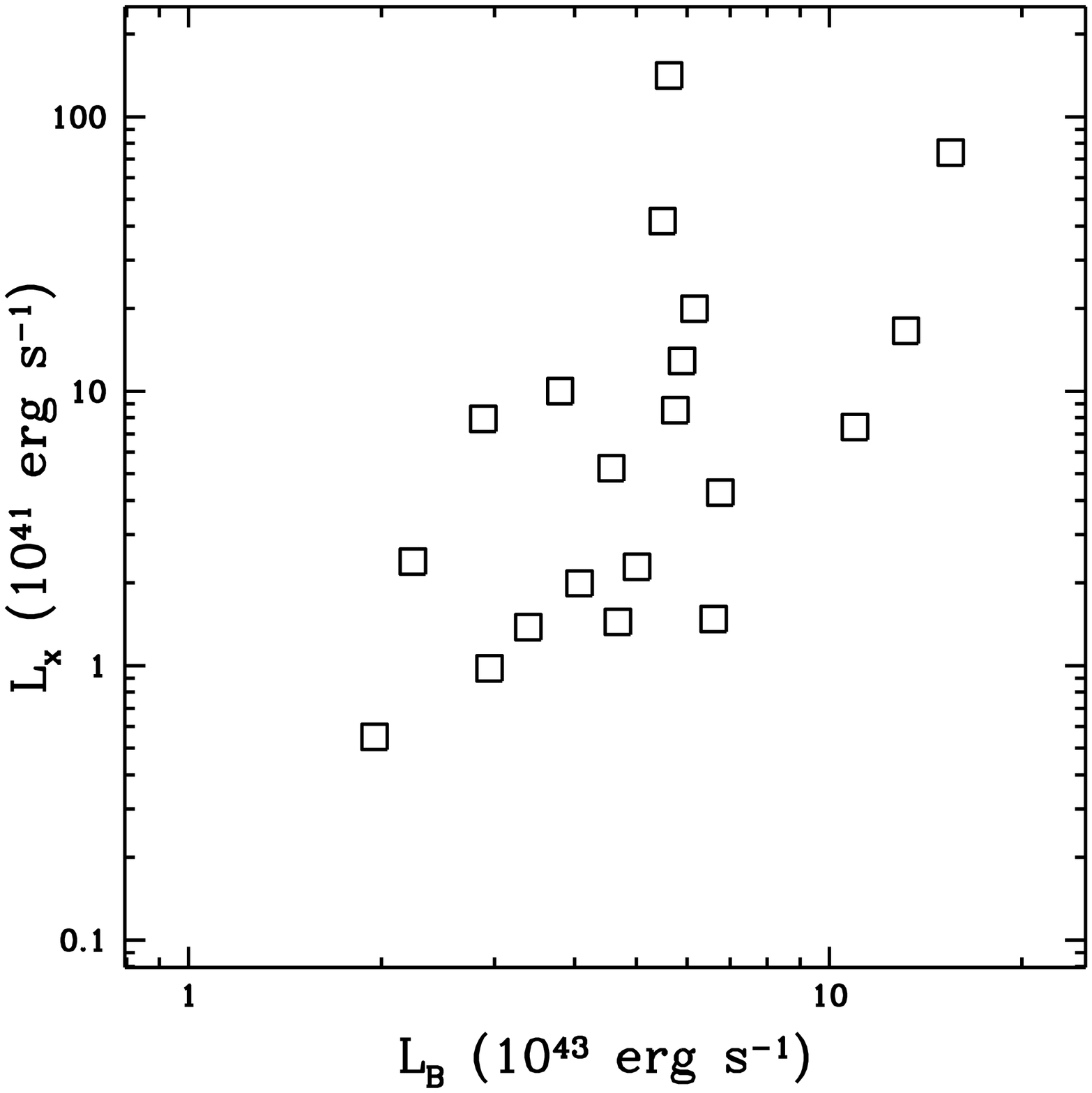,angle=0,height=0.3\textheight}}
}
\parbox{0.49\textwidth}{
\centerline{\psfig{figure=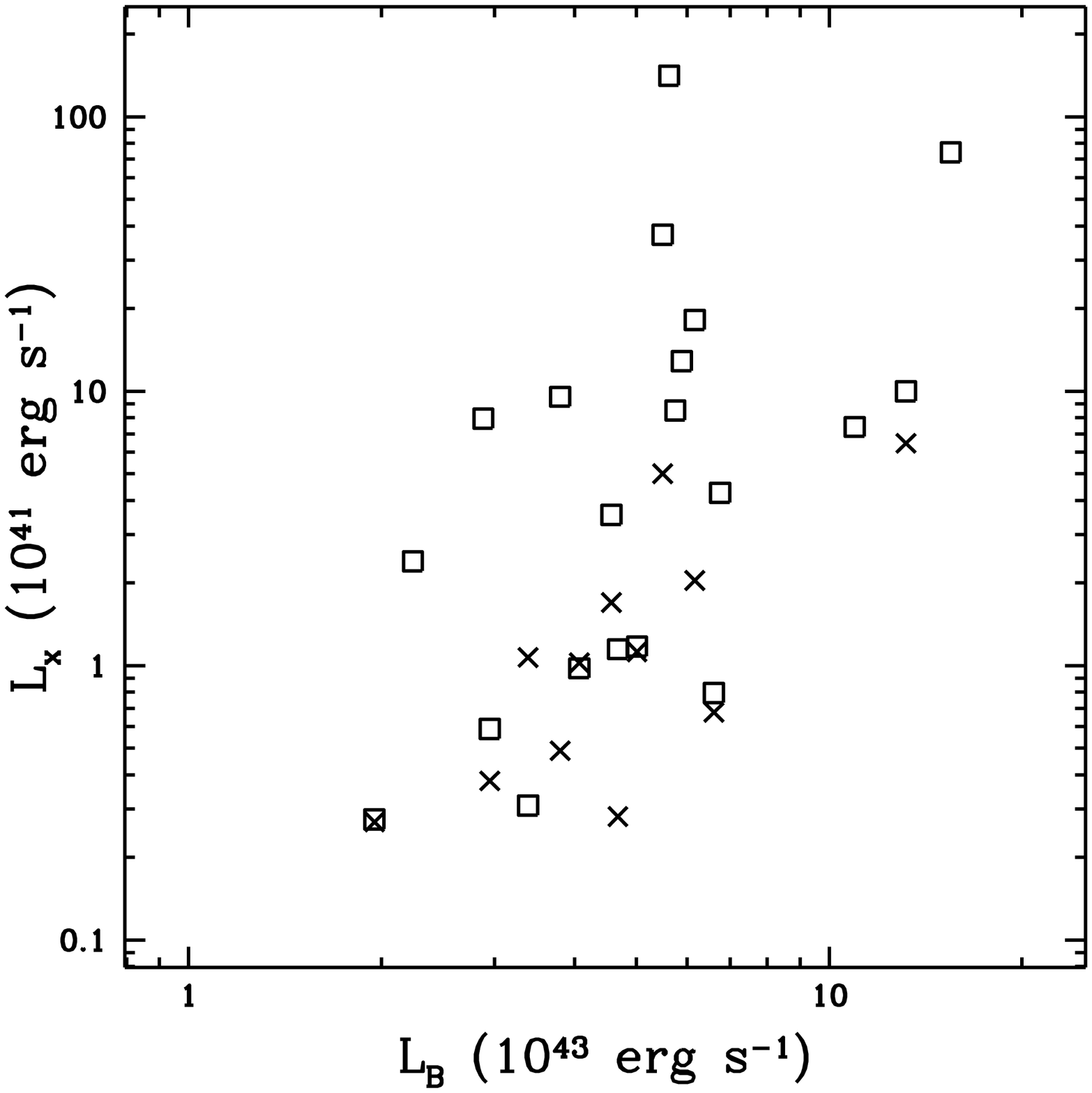,angle=0,height=0.3\textheight}}
}
\caption{\label{fig.lx} (left) The total 0.5-5 keV band X-ray
luminosity versus $L_{\rm B}$ and (right) 0.5-5 keV band luminosities of the
cold components (boxes) and hard components (crosses). }
\end{figure*}

In Table \ref{tab.lum} we give the unabsorbed luminosities of the cold
and hot components obtained from the fits of the two MEKAL models to
the galaxy spectra as well as the ratio of total X-ray luminosity,
$L_{\rm x}=L_{\rm x}^C+L_{\rm x}^H$, to $L_{\rm B}$.  (We give only
the best-fit values of $L_{\rm x}$ since the statistical uncertainties
are less than a few percent. The statistical errors on the relative
contribution of $L_{\rm x}^C$ and $L_{\rm x}^H$ are very nearly those
of $f_{\rm C}/f_{\rm H}$ in Table \ref{tab.2t}.)  Because of our procedure of
maximizing the S/N of the extraction radii (\S \ref{obs}) our
luminosities tend to be somewhat smaller than found in previous
studies which is most significant for galaxies having the lowest S/N
observations. However, the 0.5-5 keV fluxes determined in this paper
differ by $\la 25\%$ for the corresponding galaxies studied by other
authors using larger aperture sizes (Awaki et al. 1994; Matsushita et
al. 1994; Loewenstein et al. 1994; Buote \& Canizares 1997; Arimoto et
al. 1997; Matsumoto et al. 1997), and thus the luminosities in Table
\ref{tab.lum} are close to the total values.

We plot the total X-ray luminosity versus $L_{\rm B}$ in Figure
\ref{fig.lx}. As expected for the relatively high $L_{\rm x}/L_{\rm B}$ galaxies
in our sample \cite{cft}, the slope is quite steep, $d\ln L_{\rm x}/d\ln
L_{\rm B}\sim 2$, and is consistent with that predicted from steady-state
cooling flow models (Nulsen et al. 1984; Sarazin 1997). Although our
sample is too small to make a definitive statement, the slope appears
to steepen for larger $L_{\rm x}$ and $L_{\rm B}$. These galaxies include NGC 1399,
NGC 5044, and NGC 5846 for which significant emission may be due to a
diffuse intergroup medium. 

Also in Figure \ref{fig.lx} we plot the emission from the soft and
hard components separately. We take the emission of the soft
components to be given by those of the cold component with temperature
$T_{\rm C}$. If $T_{\rm H}<2$ keV then we add that emission to the soft component
as well.  The hard components are those having $T_{\rm H}\ga 5$ keV. The
slope of the soft emission appears to be slightly steeper than that of
the total X-ray emission for the galaxies with smaller $L_{\rm x}$. We can
not distinguish the slopes of the soft and hard components (though the
galaxies with measured hard components do not extend to the highest
$L_{\rm x}$ in our sample), but the hard component does appear to flatten
somewhat at the largest $L_{\rm B}$.

In any event, the slope for the hard emission is still considerably
steeper than 1 which indicates for the galaxies in our sample the hard
components are not entirely associated with discrete sources. It is
likely that with higher S/N data the temperatures of many of these
hard components will be split into subcomponents some of which will
have smaller temperatures consistent with emission from hot gas.

\begin{figure}
\centerline{\hspace{0cm}\psfig{figure=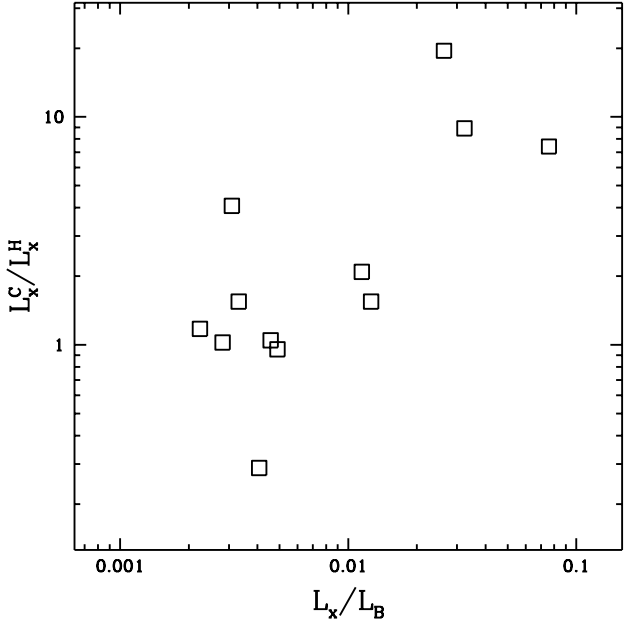,width=0.49\textwidth,angle=0}}
\caption{\label{fig.lxlb} The ratio of $L_{\rm x}^{\rm C}$ (the
luminosity of the cold component) to $L_{\rm x}^{\rm H}$ (the
luminosity of the hot component) versus $L_{\rm x}/L_{\rm B}$, where
$L_{\rm x}=L_{\rm x}^{\rm C} + L_{\rm x}^{\rm H}$.  Only galaxies with
$T_{\rm H} >\sim 5$ keV are plotted, while those with $T_{\rm H} <\sim
2$ keV (indicating another hot gas phase) would appear somewhere in
the upper right corner of the figure. All luminosities are evaluated
in the 0.5-5 keV band. }
\end{figure}

As argued in previous studies of large samples of ellipticals with the
{\it Einstein Observatory} (Canizares et al. 1987; Kim et al. 1992)
the ratio $L_{\rm x}/L_{\rm B}$ is a good indicator of when the X-ray
emission is dominated by hot gas (larger values) or by discrete
sources (smaller values). In Figure \ref{fig.lxlb} we plot the ratio
of the luminosity of the cold component to the hot component, $L_{\rm
x}^C/L_{\rm x}^H$, versus $L_{\rm x}/L_{\rm B}$. Galaxies which have
$T_{\rm H}<2$ keV are not plotted since they have no measured
component presumably due to discrete sources, but would appear
somewhere in the top right of the plot. We see an increase of $L_{\rm
x}^C/L_{\rm x}^H$ with $L_{\rm x}/L_{\rm B}$ as expected, albeit with
large scatter, and a slope slightly less than 1.  Hence, from direct
measurement of the ratio $L_{\rm x}^C/L_{\rm x}^H$ we have shown that
as $L_{\rm x}/L_{\rm B}$ increases the proportion of the total
emission due to hot gas with respect to that of a hard component
presumably due to discrete sources increases such that $d\ln (L_{\rm
x}^C/L_{\rm x}^H) / d\ln(L_{\rm x}/L_{\rm B})\sim 1$. Interestingly,
for a general slope $\alpha$ this relation implies that when $L_{\rm
x}^C/L_{\rm B}\rightarrow \infty$ we have $L_{\rm x}^H\rightarrow
L_{\rm B}/\alpha^2$ which is consistent with expectations of a
discrete component where $L_{\rm x}^{disc}\propto L_{\rm B}$.

\begin{figure}
\centerline{\hspace{0cm}\psfig{figure=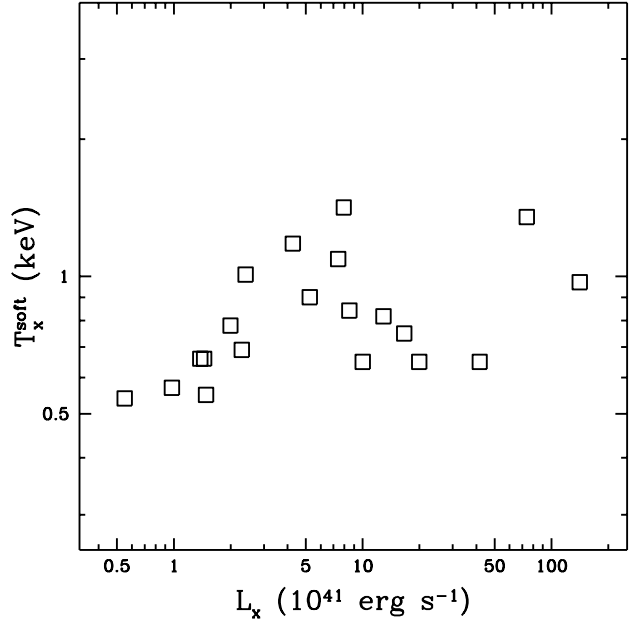,width=0.49\textwidth,angle=0}}
\caption{\label{fig.lxtx} The temperature of the soft component versus
the 0.5-5 keV total X-ray luminosity. The soft temperature is equal to
$T_{\rm C}$ for galaxies with $T_{\rm H} >\sim 5$ keV and is the
emission weighted value of $T_{\rm C}$ and $T_{\rm H}$ for galaxies
where $T_{\rm H} <\sim 2$ keV.}
\end{figure}

Finally, we plot the temperature of the soft component versus total
X-ray luminosity in Figure \ref{fig.lxtx}. This temperature is taken
to be $T_{\rm C}$ unless $T_{\rm H}<2$ keV in which case it is set to
the emission-weighted value of $T_{\rm C}$ and $T_{\rm H}$.  Over the
decade in luminosity $L_{\rm x} = (0.5-5)\times 10^{41}$ erg cm$^{-2}$
s$^{-1}$ we see the expected rise of temperature of the hot gas with
total X-ray luminosity. However, the trend disappears into scatter for
larger $L_{\rm x}$. For example, NGC 7619 and NGC 7626 have low
$T_{\rm C}$ for their X-ray luminosity. But they also have $T_{\rm
H}\sim 3$ keV and thus their hot-gas temperatures may in fact be
higher in reality.

Another explanation is given by cooling flows. We infer mass
deposition rates of $\dot{M}_{\rm gas}>1M_{\sun}$ yr$^{-1}$ for
galaxies NGC 499, NGC 4636, NGC 7619, NGC 7626, and IC 4296 which
appear to have low soft temperatures for their X-ray
luminosity. Hence, as in clusters of galaxies \cite{f94}, the
presence of cooling flows may account for much of the scatter in the
$L_{\rm x}-T_{\rm x}$ relationship.

\section{The Evidence for Excess Absorption and its
Implications}
\label{nh}

\begin{figure*}
\parbox{0.49\textwidth}{
\centerline{\psfig{figure=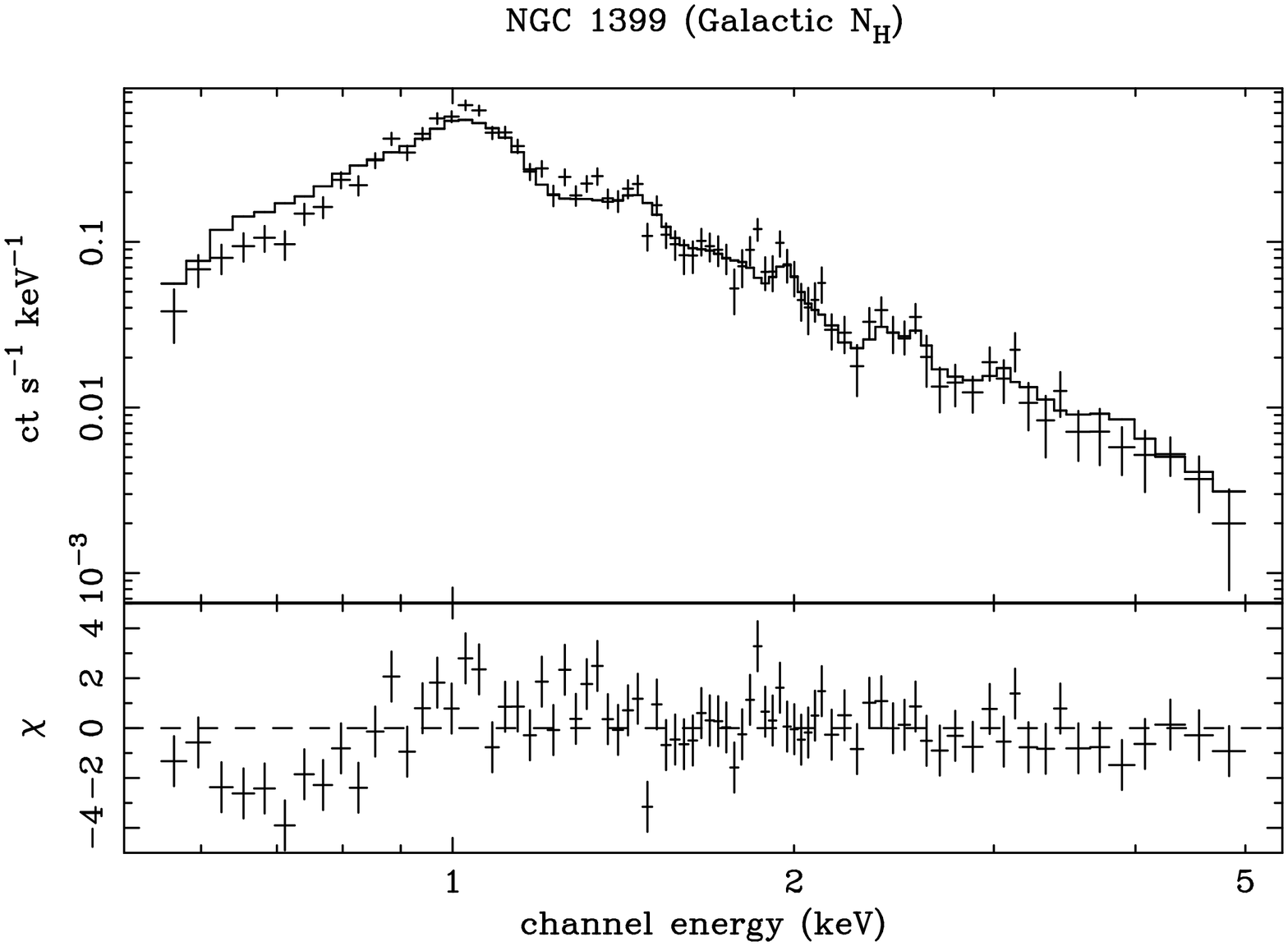,angle=-90,height=0.3\textheight}}
}
\parbox{0.49\textwidth}{
\centerline{\psfig{figure=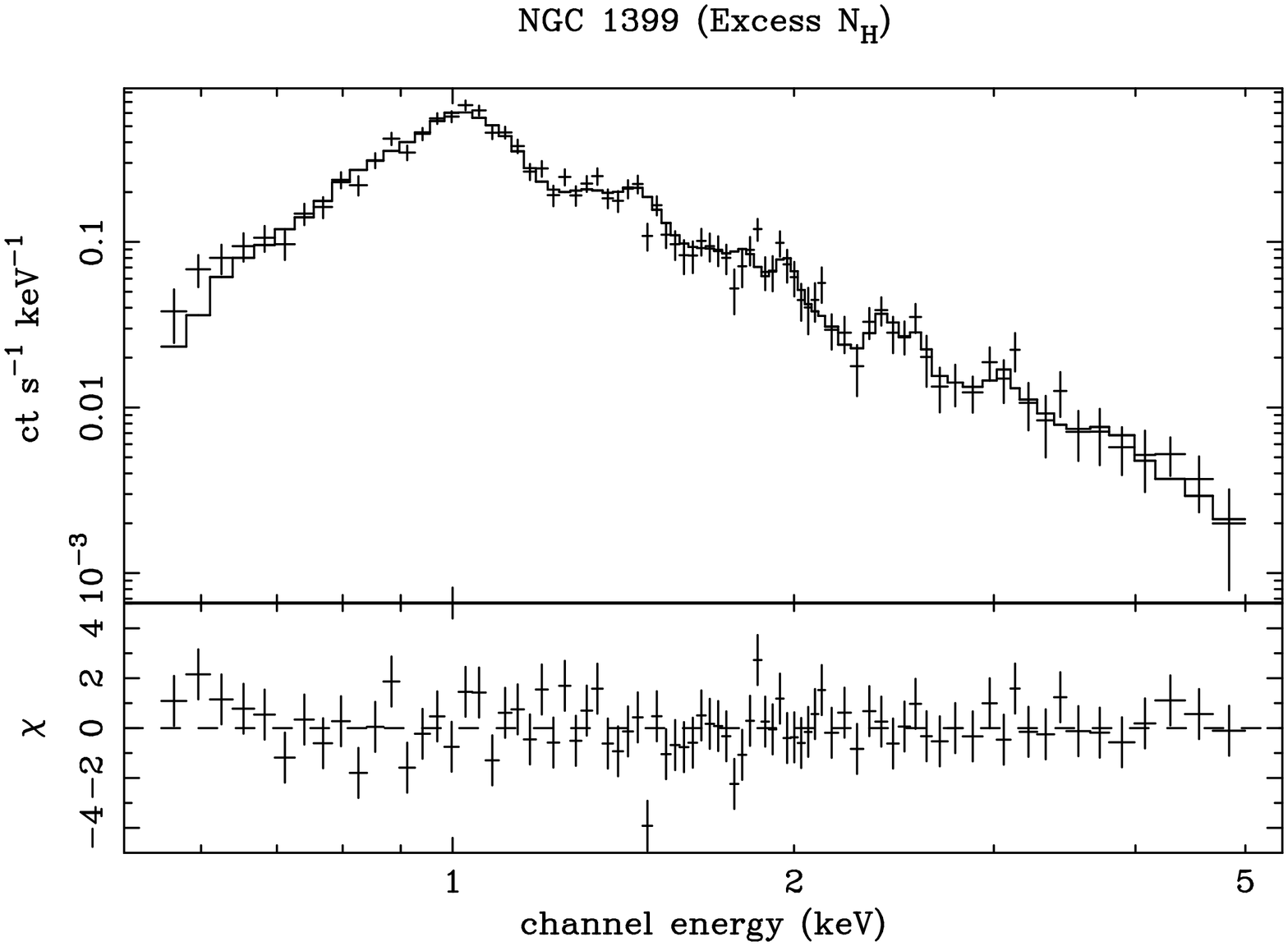,angle=-90,height=0.3\textheight}}
}
\caption{\label{fig.nh} Fits of the absorbed cooling-flow model to NGC
1399 with fixed Galactic absorption (left) and with variable
absorption (right). Only the SIS0 data is shown for sequence
80038000.}
\end{figure*}

For approximately half the galaxies in our sample the spectral fits
are significantly improved when allowing for absorption in excess of
the Galactic value. This effect is most clearly illustrated for the
galaxies where the cooling-flow model provides a substantially better
fit than a single MEKAL (or Raymond-Smith) model (see \S \ref{1t}). In
Figure \ref{fig.nh} we display the SIS0 spectrum of NGC 1399 and the
best-fit cooling-flow model obtained from jointly fitting the SIS0 and
SIS1 data of the two available observations (see Table
\ref{obs}). When fixing $N_{\rm H}$ to the Galactic value the best-fit
model predicts more emission than observed for energies $\sim 0.6-0.8$
keV: $\chi^2=575.0$ for 257 dof giving $\chi^2_{\rm red}=2.24$. These
residuals around 0.8 keV cannot be eliminated by including another
thermal component.  However, if the absorbing column is allowed to be
a free parameter then these residuals can be largely removed:
$\chi^2=341.3$ for 256 dof giving $\chi^2_{\rm red}=1.33$. The
resulting column density, $N_{\rm H}=1.8_{-0.2}^{+0.3}\times 10^{21}$
cm$^{-2}$, is over ten times the Galactic value of $N_{\rm
H}=0.13\times 10^{21}$ cm$^{-2}$ \cite{stark}.

Is this excess absorption an instrumental effect peculiar to {\sl
ASCA}? As explained in \S \ref{1t}, most previous {\sl ROSAT} studies
do not find evidence for excess absorption in early-type
galaxies. This discrepancy between the spectral shape of {\sl ROSAT}
and {\sl ASCA} below 1 keV has been observed, e.g. for the bright
Seyfert galaxies MCG-6-30-15 \cite{np} and NGC 5548 \cite{kazushi} as
well as the classic QSO 3C273 \cite{yaq}. However, a recent analysis
of MCG-6-30-15 by Orr et al. \shortcite{orr97a} finds that the $\sim
0.1-10$ keV spectrum observed by Beppo-SAX \cite{sax} agrees extremely
well with that of the {\sl ASCA} SIS. Similar agreement between
Beppo-SAX and {\sl ASCA} has also been reported for 3C273
\cite{orr97b}. Furthermore, from a simultaneous {\sl ROSAT} PSPC and
ASCA observation of the Seyfert galaxy NGC 5548 \cite{kazushi} it has
been found that the absorption determined from {\sl ASCA} agrees with
determinations from GINGA unlike the determination from {\sl
ROSAT}. Given the agreement between the spectra of {\sl ASCA}, {\sl
SAX}, and {\sl GINGA} it would appear that the excess absorption
measured by {\sl ASCA} is not the result of an instrumental effect.

We do note that the excess column density reported here, if due to a
uniform screen of the galaxy, precludes any emission being detected in
the carbon band of {\sl ROSAT}. Of course, the excess absorption is
likely to be distributed throughout the emitting gas, and the nearest
unabsorbed parts of the galaxy will give detectable radiation in that
band. (See Allen \& Fabian \shortcite{af} for simulations of this
effect for clusters of galaxies.)

We mention that Rangarajan et al. \shortcite{vijay} did find excess
absorption for NGC 1399 by exploiting the spatial information provided
by the {\sl ROSAT} PSPC. By comparing the ratios of the spectrum of
the inner region ($r\la 10$ kpc) to a best-fit unabsorbed thermal
model with the corresponding ratio of the spectrum of an outer annulus
($r\sim 100-150$ kpc) to its best-fit unabsorbed thermal model they
clearly showed the diminution of photons with energies $\la 1$ keV in
the inner region (see their Figure 5).  However, consistent with the
general {\sl ASCA}-{\sl ROSAT} discrepancy stated above, excess
absorption is not required when fitting absorbed (single-component)
thermal models to the data \cite{jones}.

The excess $N_{\rm H}$ we have derived from the spectral fits is only
the simplest result assuming a uniform absorber which reduces the
X-ray emission according to an exponential form depending only on the
energy. In this approximation the excess columns $N_{\rm H}\approx
(1-3)\times 10^{21}$ cm$^{-2}$ imply masses in atomic hydrogen,
$M_{\rm abs}\approx 5\times 10^9 M_{\sun}$, when spread uniformly over
a 10 kpc radius. This level of absorber mass is very similar to the
typical mass of the hot X-ray emitting gas in early-type galaxies, but
$M_{\rm abs}$ is typically a factor of 10 larger than the mass
inferred from HI observations (e.g. Bregman, Roberts, \& Giovanelli
1988; Bregman \& Roberts 1990). Similar disagreement is found for an
absorption feature in the field of NGC 4472 by Irwin \& Sarazin
\shortcite{is}. However, these discrepancies with HI observations may
be irrelevant since the absorber could be composed of molecular gas
(e.g. Ferland, Fabian, \& Johnstone 1994). The mass of molecular gas
can be measured with CO observations, but is typically factors of a
few smaller than expected under the above simple assumptions (e.g.
Irwin et al. 1997). However, excess absorbing mass above that
predicted by the standard CO/$H_2$ conversion has been reported from
158 micron observations of the irregular galaxy IC 10 \cite{madden}.
It should be emphasized that the uniform absorber model is certainly
over-simplified and, e.g. a partial covering fraction and metallicity
dependence could very well be required.

The cold gas implied by the excess absorption is a natural consequence
of the multi-phase cooling-flow scenario (Nulsen 1986; Thomas et
al. 1987; White \& Sarazin 1987) wherein mass drops out of the flow at
all radii where the cooling time is less than the age of the galaxy
($t_{\rm age}$). In particular, for clusters of galaxies Allen \&
Fabian \shortcite{af} have demonstrated using color profiles of {\sl
ROSAT} PSPC data that clusters with $t_{\rm cool}\ll t_{\rm age}$ have
intrinsic absorbing material which is preferentially distributed in
the core and of a quantity that is consistent with the amount of mass
deposited by the cooling flow over $t_{\rm age}$.

Although the situation in early-type galaxies is likely complicated to
some extent by heat sources, the success of the simple cooling-flow
model over the single MEKAL model for the brightest galaxies and the
associated required excess absorption (see \S \ref{1t}) suggests that
multi-phase cooling flows are operating in these galaxies.  This is
further supported by the fact that these galaxies when fitted with
two-temperatures indicate that they have a multi-phase medium (see \S
\ref{2t}); i.e. one component may represent the hot ambient phase and
the other the emission from the distributed mass drop-out
component. In this scenario, the drop-out term should consist of a
continuum of phases which will become evident when probed by detectors
with better spectral resolution.

\section{Conclusions}
\label{conc}

We present spectral analysis of {\sl ASCA} data of 20 bright, mostly
high $L_{\rm x}/L_{\rm B}$, early-type galaxies (17 E, 3 S0) which is
the largest sample of its kind reported to date. Fitting a single,
absorbed, thin thermal model (MEKAL or Raymond-Smith) to the spectra
fails to provide acceptable fits (e.g. $\chi^2_{\rm red} > 1.5$) for
14 galaxies. Excess absorption significantly improves the fits for 10
of the galaxies. The abundances of the single-temperature models are
very sub-solar, $\langle Z\rangle = 0.19\pm 0.12$ $Z_{\sun}$, in
agreement with those reported by previous {\sl ROSAT} and {\sl ASCA}
studies.

Fitting instead an absorbed cooling-flow model yields fits of
comparable quality to the MEKAL model for 11 galaxies but is a
significantly better fit for the other 9. The improvement in the fits
is dramatic for NGC 4472, NGC 5044, and IC 4296. The abundances
obtained from the cooling-flow model, $\langle Z\rangle = 0.6\pm 0.5$
$Z_{\sun}$, significantly exceed those obtained with the MEKAL and
Raymond-Smith models.

We empirically tested the reliability of the MEKAL and Raymond-Smith
plasma codes in the Fe-L band. In order to lessen the influence of
other temperature components we restricted our analysis to 0.5-2 keV
and performed fits over only the Fe-L band (taken to be 0.7-1.2 keV)
and over the band excluding Fe-L (0.5-0.7 and 1.2-2 keV). We did not
find large systematic trends expected if the plasma codes predict too
much or too little Fe-L emission for a given temperature and
metallicity. However, unlike the MEKAL model, the Raymond-Smith
model did give some evidence for these systematic trends which, as
expected, indicates that the Fe-L region of Raymond-Smith is not as
well modeled as with MEKAL. (The reasonable agreement of the
abundances obtained by MEKAL and Raymond-Smith shows that such errors
are not catastrophic.)

Fitting two-temperature models (with MEKAL or Raymond-Smith models)
significantly improved the quality of the fit in 16 out of 20 galaxies
and gave formally acceptable values of $\chi^2_{\rm red}\sim 1.0$. (We
found that care needed to be exercised with XSPEC to achieve the
global $\chi^2$ minima for the lowest S/N galaxies in the sample.)
The galaxies having the highest S/N spectra in our sample (which
coincidentally have the largest $L_{\rm x}/L_{\rm B}$) generally have
a temperature for the cold component, $T_{\rm C}\sim 0.5-1$ keV, and
for the hot component, $T_{\rm H}\sim 1-2$ keV; i.e. both components
are consistent with emission from hot gas\footnote{Recently
Loewenstein \& Mushotzky \shortcite{lmushy} claim that ratios of the
H-like to He-like Si lines in, e.g. NGC 1399 are only consistent with
an isothermal gas with temperature near 1 keV and inconsistent with
the multi-phase solution we find in \S \ref{2t}. We have examined such
line ratios and find that the line ratios of the single-temperature
models in \S \ref{1t} are very similar to those of the two-temperature
models in \S \ref{2t} (i.e. within $\sim 15\%$). However, caution must
be exercised when computing these line ratios directly from the data
because of the calibration errors due to the optical constants in the
mirror around 2 keV \cite{gendreau}. Since the key H-like Si line
appears at 2.0 keV, and for all of the galaxies in our sample this
line is very noisy (e.g. Figure \ref{fig.nh}), and its intensity
depends sensitively on the assumed continuum, we believe that any
reliable constraints on the 2.0 keV H-like Si line must await
corrected response matrices.}.  The lower S/N galaxies (which
coincidentally have the smallest $L_{\rm x}/L_{\rm B}$) typically have
$T_{\rm C}\sim 0.5-1$ keV but $T_{\rm H}\ga 5$ keV. The temperature of
this hot component is consistent with emission from discrete
sources. The flux of this hot component is generally smaller than that
of the cold component.

In stark contrast to the single-temperature models, the abundances
obtained from the two-temperature fits are approximately solar;
$\langle Z\rangle = 0.9\pm 0.7$ $Z_{\sun}$ for two MEKAL models and
$\langle Z\rangle = 0.7\pm 0.6$ $Z_{\sun}$ for two Raymond-Smith
models. The difference between the abundances determined from
single-temperature and two-temperature models is not due to increased
errors on the parameters but is an artifact of fitting intrinsically
multi-temperature spectra with single-temperature models \cite{BC94}.

Our abundances disagree with the two-component models fit to {\it
ASCA} spectra of several of the galaxies in our sample by Matsumoto et
al. \shortcite{mat}. We attribute the discrepancy to the different
modeling procedures used and to the higher S/N data in our study owing
to generally smaller extraction radii for the spectra.  Possible
abundance gradients do not account for the discrepancy because we
obtain approximately solar abundances even for some of the large S/N
galaxies in our sample which have large extraction radii (e.g. NGC
1399, NGC 4472).  The approximately solar abundances we find are
consistent with multi-phase models for the evolution of the hot gas in
ellipticals (e.g. Fujita et al. 1997).  However, though considerably
larger than obtained in previous studies, these nearly solar
abundances are still substantially less than the super-solar
abundances predicted by the standard enrichment theories (e.g. Ciotti
et al. 1991; also see Arimoto et al. 1997).

Two-component models where one component is a cooling-flow model and
the other a MEKAL give fits of comparable quality to fits of two MEKAL
models. Since the flux is shared by the two components, the inferred
mass deposition rates are smaller than when only a cooling-flow model
is fit to the spectra.

It should be emphasized that a radial temperature gradient in a
single-phase gas unlikely accounts for the temperatures seen in the
large $L_{\rm x}/L_{\rm B}$ galaxies. For example, NGC 4472 has
roughly an equal contribution from phases with temperatures $T_{\rm
C}=0.76$ keV and $T_{\rm H}=1.48$ keV, but the temperature profile
measured with the {\sl ROSAT} PSPC varies from only $\sim 0.75-1.1$
keV over $r\le 4\arcmin$ \cite{is}. A plausible physical explanation
of the two phases is offered by a multi-phase cooling flow (Nulsen
1986; Thomas et al. 1987; White \& Sarazin 1987) where one component
represents the hot ambient phase and the other represents the emission
from the distributed mass drop-out component. In this scenario, the
drop-out term should consist of a continuum of phases which should
become evident when probed by detectors with better spectral
resolution.

We compared $L_{\rm B}$ to the luminosities of the hot and cold
components determined by fits of two MEKAL models. Both the
luminosities of the soft and hard components increase with $L_{\rm B}$
with a slope approximately 2, although there is some weak evidence
that the cold component has a steeper slope than the hot component.
For galaxies with $T_{\rm H}\ga 5$ keV we find that $L_{\rm
x}^C/L_{\rm x}^H \sim L_{\rm x}/L_{\rm B}$; i.e. there is a clear
increase in the ratio of the luminosity of the cold component to the
hot component as a function of $L_{\rm x}/L_{\rm B}$ indicating that
$L_{\rm x}/L_{\rm B}$ is indeed a good indicator for the relative
contribution of hot gas and discrete sources to the total X-ray
emission. (A recent study of NGC 3923 by Buote \& Canizares (1998)
finds that the allowed contribution to the {\sl ROSAT} X-ray emission
from a population of discrete sources with $L_{\rm x}\propto L_{\rm
optical}$ is significantly less than indicated by $L_{\rm x}^C/L_{\rm
x}^H$ measured in this paper if we associate $L_{\rm x}^H \propto
L_{\rm optical}$. This would support the conjecture that a significant
fraction of the hard components measured for the low S/N galaxies is
due to other phases in the hot gas.)

Therefore, our investigation shows that the very sub-solar abundances
obtained in previous studies are not the result of failures in the
plasma codes (as proposed by Arimoto et al. 1997). Rather, fitting
multi-temperature models, without restricting the form of the models
to a universal model for discrete sources, and setting extraction
radii so that S/N is maximized, gives metal abundances near solar for
both MEKAL and Raymond-Smith models.

Moreover, fits to the X-ray spectra of half of the galaxies in our
sample are significantly improved when allowing for absorption in
excess of the Galactic value. Accurate determination of the spatial
distributions of the absorbing components and the various phases of
the hot gas must await better quality data as must determination of
the nature of the hard components found in the lower S/N galaxies in
our sample. The next generation of X-ray satellites soon to be flown
will have the capabilities to make significant progress in these
areas.

\section*{Acknowledgments}

We acknowledge helpful discussions with S. Allen and K. Iwasawa on the
reduction of {\sl ASCA} data and K. Gendreau on {\sl ASCA}
calibration.  This research has made use of (1) ASCA data obtained
from the High Energy Astrophysics Science Archive Research Center
(HEASARC), provided by NASA's Goddard Space Flight Center and (2) the
NASA/IPAC Extragalactic Database (NED) which is operated by the Jet
Propulsion Laboratory, California Institute of Technology, under
contract with the National Aeronautics and Space Administration.


\begin{thebibliography}{}
\bibitem[\protect\citename{Allen \& Fabian }1997]{af}
Allen S. W., Fabian A. C., 1997, MNRAS, 286, 583
\bibitem[\protect\citename{Anders \& Grevesse }1989]{ag}
Anders E., Grevesse N., 1989, Geochimica et Cosmochimica Acta,
53, 197
\bibitem[\protect\citename{Arimoto et al. }1997]{arimoto}
Arimoto N., Matsushita K., Ishimaru Y., Ohashi T., Renzini A., 1997,
ApJ, 477, 128
\bibitem[\protect\citename{Arnaud }1996]{xspec} 
Arnaud K., 1996, in Jacoby G. and Barnes J., eds., 
Astronomical Data Analysis Software and Systems V, ASP Conf. Series
volume 101, p17 
\bibitem[\protect\citename{ASCA Guide }1997]{abc}
ASCA Guest Observer Facility, The ASCA Data Reduction Guide,
Laboratory for High Energy Astrophysics (Greenbelt: NASA/GSFC)
\bibitem[\protect\citename{Awaki et al. }1994]{awaki}
Awaki H., et al., 1994, PASJ, 46, L65
\bibitem[\protect\citename{Baluci\'{n}ska-Church \& McCammon
}1992]{phabs} 
Baluci\'{n}ska-Church M., McCammon D., 1992, 400, 699
\bibitem[\protect\citename{Bregman \& Roberts }1990]{br}
Bregman J. N., Roberts S., 1990, ApJ, 362, 828
\bibitem[\protect\citename{Bregman et al. }1988]{brg}
Bregman J. N., Roberts S., Giovanelli R., 1988, ApJ, 330, L93
\bibitem[\protect\citename{Buote \& Canizares }1994]{BC94}
Buote D. A., Canizares C. R., 1994, ApJ, 427, 86
\bibitem[\protect\citename{Buote \& Canizares }1997]{BC97}
Buote D. A., Canizares C. R., 1997, ApJ, 474, 650
\bibitem[\protect\citename{Buote \& Canizares }1998]{BC98}
Buote D. A., Canizares C. R., 1998, MNRAS, submitted
\bibitem[\protect\citename{Canizares et al. }1987]{cft}
Canizares C. R., Fabbiano G., Trinchieri G., 1987, ApJ, 312, 503
\bibitem[\protect\citename{Ciotti et al. }1991]{ciotti}
Ciotti L., D'Ercole A., Pellegrini S., Renzini A., 1991, ApJ, 376, 380
\bibitem[\protect\citename{Dalle Ore et al. }1987]{dalle}
Dalle Ore C., Faber S. M., Jes\'{u}s J., Stoughton R., 1991, ApJ, 366,
38
\bibitem[\protect\citename{David et al. }1994]{david}
David L. P., Jones C., Forman W., Daines S., 1994, ApJ, 428, 544
\bibitem[\protect\citename{Davis \& White }1996]{dw}
Davis D. S., White III R. E., 1996, ApJ, 470, L35
\bibitem[\protect\citename{de Vaucouleurs et al. }1991]{rc3}
de Vaucouleurs G., de Vaucouleurs A., Corwin H.G., Buta R. J., Paturel
G., Fouqu\'{e} P., 1991, Third Reference Catalogue of Bright Galaxies
(Austin: Univ. Texas Press) (RC3)
\bibitem[\protect\citename{Eskridge et al. }1995]{efk}
Eskridge P. B., Fabbiano G., Kim D.-W., 1995, ApJS, 97, 141
\bibitem[\protect\citename{Fabbiano et al. }1994]{fkt}
Fabbiano G., Kim D.-W., Trinchieri G., 1992, ApJS, 80, 531
\bibitem[\protect\citename{Faber et al. }1989]{7s}
Faber S. M., Wegner G., Burstein D., Davies R. L., Dressler A.,
Lynden-Bell D., Terlevich R. J., 1989, ApJS, 69, 763
\bibitem[\protect\citename{Fabian }1994]{acf}
Fabian A. C., 1994, ARA\&A, 32, 277
\bibitem[\protect\citename{Fabian et al. }1994]{f94}
Fabian A. C., Crawford C. S., Edge A. C., Mushotzky R. F., 1994,
MNRAS, 267, 779
\bibitem[\protect\citename{Ferland et al. }1994]{ferland}
Ferland G. J.,  Fabian A. C., Johnstone R. M., 1994, MNRAS, 266, 399
\bibitem[\protect\citename{Forman et al. }1985]{fjt}
Forman W., Jones C., Tucker W., ApJ, 1985, 293, 102
\bibitem[\protect\citename{Forman et al. }1979]{f79} 
Forman W., Schwarz J., Jones C., Liller W., Fabian, A. 1979, ApJ, 234,
L27
\bibitem[\protect\citename{Forman et al. }1993]{forman}
Forman W., Jones C., David L., Franx M., Makishima K., Ohashi T.,
1993, ApJ, 418, L55
\bibitem[\protect\citename{Fujita et al. }1997]{ffo}
Fujita Y., Fukumoto J., Okoshi K., 1997, ApJ, 488, 585
\bibitem[\protect\citename{Fukazawa et al. }1996]{fuk}
Fukazawa Y., et al., 1996, PASJ, 48, 395 
\bibitem[\protect\citename{Garcia }1993]{garcia}
Garcia A. M., 1993, A\&AS, 100, 47
\bibitem[\protect\citename{Gendreau \& Yaqoob }1997]{gendreau}
Gendreau K., Yaqoob T., 1997, ASCA XRT Calibration Issues, in ASCA News
vol. 5, (NASA/GSFC: Greenbelt)
\bibitem[\protect\citename{Idesawa et al. }1997]{idesawa}
Idesawa E., et al., 1997, Calibration of Temporal and Spatial
Variations of the GIS Gain,
ftp://legacy.gsfc.nasa.gov/asca/gis\_information/gain.ps/gz 
\bibitem[\protect\citename{Irwin \& Sarazin }1996]{is}
Irwin J. A., Sarazin C. L., 1996, ApJ, 471, 683
\bibitem[\protect\citename{Irwin et al. }1997]{ifs}
Irwin J. A., Frayer D. T., Sarazin C. L., 1997, in Soker N. ed.,
Galactic and Cluster Cooling Flows, ASP Conference Series vol. 115,
(ASP: San Francisco), 75
\bibitem[\protect\citename{Ishisaki et al. }1997]{ishisaki}
Ishisaki Y., et al., 1997,  Reproducibility of the GIS Non--X-Ray
Background, in ASCA News vol. 5, (NASA/GSFC: Greenbelt)
\bibitem[\protect\citename{Iwasawa et al. }1997a]{kazu}
Iwasawa K., White D. A., Fabian A. C., 1997a, MNRAS, submitted
\bibitem[\protect\citename{Iwasawa et al. }1997b]{kazushi}
Iwasawa K., et al., 1997b, in preparation
\bibitem[\protect\citename{Johnstone et al. }1992]{rjcool}
Johnstone R. M., Fabian A. C., Edge A. C., Thomas P. A., 1992, MNRAS,
255, 431
\bibitem[\protect\citename{Jones et al. }1997]{jones}
Jones C., Stern C., Forman W., Breen J., David L., Tucker W., Franx
M., 1997, ApJ, 482, 143
\bibitem[\protect\citename{Kaastra \& Mewe }1992]{km}
Kaastra J. S., Mewe R., 1993, A\&AS, 97, 443
\bibitem[\protect\citename{Kim \& Fabbiano }1995]{kf}
Kim D.-W., Fabbiano G., 1995, ApJ, 441, 182
\bibitem[\protect\citename{Kim et al. }1992]{kft}
Kim D.-W., Fabbiano G., Trinchieri G., 1992, ApJ, 393, 134
\bibitem[\protect\citename{Liedahl et. al }1995]{mekal}
Liedahl D. A., Osterheld A. L., Goldstein W. H., 1995, ApJ, 438, L115
\bibitem[\protect\citename{Loewenstein \& Mathews }1991]{lm}
Loewenstein M., Mathews W. G., 1991, ApJ, 373, 445
\bibitem[\protect\citename{Loewenstein \& Mushotzky }1997]{lmushy}
Loewenstein M., Mushotzky R. F., 1997, Proceedings of IAU Symposium
187 on Cosmic Chemical Evolution, in press (astro-ph/9710339)
\bibitem[\protect\citename{Loewenstein et. al }1994]{loew}
Loewenstein M., et al., 1994, ApJ, 436, L75
\bibitem[\protect\citename{Madden et. al }1997]{madden}
Madden S. C., Poglitsch A., Geis N., Stacey G. J., Townes C. H., 1997,
ApJ, 483, 200
\bibitem[\protect\citename{Matsumoto et. al }1997]{mat}
Matsumoto H., Koyama K., Awaki H., Tsuru T., Loewenstein M.,
Matsushita K., 1997, ApJ, 482, 133
\bibitem[\protect\citename{Matsushita et. al }1994]{matsu}
Matsushita K., et al., 1994, 436, L41
\bibitem[\protect\citename{Mewe et al. }1985]{mewe}
Mewe R., Gronenschild E. H. B. M., van den Oord G. H. J., 1985, A\&AS,
62, 197
\bibitem[\protect\citename{Nandra \& Pounds }1992]{np}
Nandra K., Pounds K. A., 1992, Nature, 359, 215
\bibitem[\protect\citename{Nulsen et al. }1986]{pejn}
Nulsen P. E. J., 1986, MNRAS, 221, 377
\bibitem[\protect\citename{Nulsen et al. }1984]{nsf}
Nulsen P. E. J., Stewart G. C., Fabian A. C., 1984, MNRAS, 208, 185
\bibitem[\protect\citename{Orr et al. }1997a]{orr97a}
Orr A., Molendi S., Fiore F., Grandi P., Parmar A. N., Owens A., 1997a,
A\&A, in press (astro-ph/9706133)
\bibitem[\protect\citename{Orr et al. }1997b]{orr97b}
Orr A., et al., 1997b, in The Active X-Ray Sky: Results from Beppo-SAX
and Rossi XTE, held 21-24 October in Rome
\bibitem[\protect\citename{Parmar et al. }1997]{sax}
Parmar A., et al., 1997, A\&AS, 122, 309
\bibitem[\protect\citename{Rangarajan et al. }1995]{vijay}
Rangarajan F. V. N., Fabian A. C., Forman W. R., Jones C., 1995,
MNRAS, 272, 665
\bibitem[\protect\citename{Raymond \& Smith }1977]{rs}
Raymond J. C., Smith B. W., 1977, ApJS, 35, 419
\bibitem[\protect\citename{Sarazin }1997]{sarazin} Sarazin C. L.,
1997, in Arnaboldi M., Da Costa G. S., and P. Saha, eds., The Nature
of Elliptical Galaxies, Proceedings of the Second Stromlo Symposium,
in press
\bibitem[\protect\citename{Stark et al. }1992]{stark}
Stark A. A., Gammie C. F., Wilson R. W., Bally J., Linke R. A., Heiles
C., Hurwitz M. 1992, ApJS, 79, 77 
\bibitem[\protect\citename{Tanaka et al. }1994]{tanaka}
Tanaka Y., Inoue H., \& Holt S. S., 1994, PASJ, 46,
L37
\bibitem[\protect\citename{Thomas et al. }1986]{thomas}
Thomas P. A., Fabian A. C., Arnaud K. A., Forman W., Jones C., 1986,
MNRAS, 222, 655
\bibitem[\protect\citename{Thomas et al. }1987]{tfn}
Thomas P. A., Fabian A. C., Nulsen P. E. J., 1987, MNRAS, 228, 973
\bibitem[\protect\citename{Trinchieri et al. }1996]{tfk}
Trinchieri G., Fabbiano G., Kim D.-W., 1996, A\&A, 318, 361
\bibitem[\protect\citename{Trinchieri et al. }1994]{trin}
Trinchieri G., Kim D.-W., Fabbiano G.,  Canizares C., 1994, ApJ, 428,
555
\bibitem[\protect\citename{White et al. }1991]{w91}
White D. A., Fabian A. C., Johnstone R. M., Mushotzky R. F., Arnaud
K. A., 1991, MNRAS, 252, 72
\bibitem[\protect\citename{White \& Sarazin }1987]{ws87}
White R. E. III, Sarazin C. L., 1987, ApJ, 318, 621
\bibitem[\protect\citename{Yaqoob et al. }1994]{yaq}
Yaqoob T., et al., 1994, PASJ, 46, L49

\end{thebibliography}
\end{document}